\documentclass[12pt, a4paper]{article}
\pdfoutput=1

\usepackage{amsmath,amssymb,exscale,xcolor}
\usepackage{psfrag,graphicx}
\usepackage{enumerate}
\usepackage[colorlinks=True, citecolor=blue, linkcolor=blue, urlcolor=black]{hyperref}
\input epsf

\newcommand{\be}{\begin{equation}}
\newcommand{\ee}{\end{equation}}
\newcommand{\bea}{\begin{eqnarray}}
\newcommand{\eea}{\end{eqnarray}}

\newcommand{\eq}[1]{Eq.~(\ref{#1})}
\newcommand{\zir}{z_{\rm IR}}
\newcommand{\zirc}{z_{\rm IR}^c}
\newcommand{\zircp}{z_{\rm IR}^{c\, \prime}}
\newcommand{\zuv}{z_{\rm UV}}
\newcommand{\f}{\frac}
\newcommand{\diag}{\operatorname{diag}}
\newcommand{\pd}{\hat\phi_d}
\newcommand{\ef}{t}

\def\gappeq{\mathrel{\rlap {\raise.5ex\hbox{$>$}}
{\lower.5ex\hbox{$\sim$}}}}
\def\lappeq{\mathrel{\rlap{\raise.5ex\hbox{$<$}}
{\lower.5ex\hbox{$\sim$}}}}
\newcommand{\Tr}{\mathop{\rm Tr}}
\def\I{\rm 1\kern-.24em l}  

\begin{document}
\pagestyle{empty}
\begin{flushright}
May 2019
\end{flushright}
\vspace*{5mm}

\begin{center}
\vspace{1.cm}
{\Large\bf
Holographic  conformal transition and  light scalars}\\
\vspace{1.5cm}
{\large Alex Pomarol$^{a,b}$, Oriol Pujolas$^a$ and Lindber Salas$^a$}\\
\vspace{.6cm}
$^a${\it {IFAE and BIST,  Universitat Aut{\`o}noma de Barcelona,
08193 Bellaterra, Barcelona}}\\
$^b${\it {Dept. de F\'isica,  Universitat Aut{\`o}noma de Barcelona,
08193 Bellaterra, Barcelona}}\\
\vspace{.4cm}
\end{center}

\vspace{1cm}
\begin{abstract}
We present an holographic approach to strongly-coupled theories close to the conformal to non-conformal  transition, 
trying to understand the presence of light scalars as recent lattice simulations seem to suggest.
We find that the dilaton is always the lightest resonance, although  not parametrically lighter than the others.
We provide a simple analytic formula for the dilaton mass that allows us to understand this behavior. 
The  pattern of the meson mass spectrum, as  we get close to  the conformal transition, is  
  found  to be quite similar to that  in  lattice simulations.  We  provide
    further predictions from holography   that can be checked in the future.
These five-dimensional models can also  implement new   solutions to the hierarchy problem, 
having implications for   searches at the LHC and cosmology.
\end{abstract}

\vfill
\eject
\pagestyle{empty}
\setcounter{page}{1}
\setcounter{footnote}{0}
\pagestyle{plain}
%

\section{Introduction}

Theories close to being  conformally  invariant  are of utmost interest as 
they can generate large hierarchies of scales that can be useful in particle physics and cosmology. 
This motivates the  understanding  of how theories behave at the critical point at which, by varying the parameters of the theory, 
we pass from a conformal regime  to a  non-conformal one.

This is especially interesting in  strongly-coupled theories as they can give rise to non-trivial dynamics.
An example is    QCD  where by increasing  the number of flavors $N_F$,   the theory  is expected to 
  become conformally invariant at some critical value $N_F=N_F^{crit}$.
It is unclear where this exactly happens, but lattice simulations suggest that this could be around $N_F^{crit}\sim 10$.
For $N_F\geq N_F^{crit}$, QCD becomes a conformal field theory (CFT) till reaching $N_F=\frac{11}{2}N_c$, where $N_c$ is the number of colors ($N_c=3$ for real QCD), at which the theory reaches the Banks-Zaks fixed point, becoming IR free for $N_F>\frac{11}{2}N_c$. The region $N_F^{crit}\leq N_F\leq \frac{11}{2}N_c$ is called the conformal window.

Recent lattice simulations suggest that, contrary to real QCD, theories close to the conformal transition
have as the lightest resonance a $0^{++}$ state (apart, of course, from the Goldstone bosons, the pions)
\cite{Aoki:2016wnc,Brower:2015owo}. 
It is unclear  what is the origin of the lightness of this state. 
Some arguments suggest that this could be a dilaton, the Goldstone associated to the breaking of scale invariance.
If this is the case, it would be interesting to know whether
in the large-$N_c$ limit, where  $N_F^{crit}/N_c\equiv x_{crit}$  becomes a continuous parameter,
the dilaton  mass tends to zero as we approach  the critical point from below $N_F/N_c\to x_{crit}$.

 In  this article we would like to analyze 
  the physics of  conformal transitions using holography.
We will follow Ref.~\cite{Kaplan:2009kr} that argued that  the exit of the conformal window of large-$N_c$ QCD  
occurs when  the IR fixed point disappears by merging with a UV fixed point.
Close to the conformal edge the theory contains a marginal operator  ${\cal O}_g$
whose  dimension gets a small imaginary part when  conformal invariance is lost
(see next section for details).
Assuming that this is the case, the AdS/CFT correspondence \cite{adscft} can provide 
a simple realization of this idea as  a complex operator dimension
matches to  a scalar  having  a  mass below  
 the Breitenlohner-Freedman (BF) bound   $M^2_\Phi=-4/L^2$.
 When this happens, the scalar becomes tachyonic and gets a non-zero profile
that results into a departure from the  Anti-de-Sitter (AdS) geometry \cite{Kaplan:2009kr}.

The presence of a   marginal operator ${\cal O}_g$ in the model could suggest the presence of a light dilaton,
along the lines of Refs.~\cite{GW,Megias:2014iwa}. The argument goes as follows.
The dilaton potential can be written as
\be
V_{\rm eff}(\phi_d)=\lambda_{\rm eff}(\phi_d)\,\phi_d^4\,,
\label{potdil}
\ee
such that, when a minimum exists, leads to a   dilaton mass given by
\be
\frac{m^2_{\phi_d}}{\langle \phi_d\rangle^2}=\beta_{\lambda_{\rm eff}} (4+\beta'_{\lambda_{\rm eff}})\,,
\label{intromass}
\ee
where $\beta_{\lambda_{\rm eff}}={d\lambda_{\rm eff}}/{d\ln\phi_d}$ and 
 $\beta'_{\lambda_{\rm eff}}=d\beta_{\lambda_{\rm eff}}/d\lambda_{\rm eff}$.
A nonzero $\beta_{\lambda_{\rm eff}}$
arises only from an explicit breaking of scale invariance.
When this latter comes only from $g\,{\cal O}_g\in {\cal L}$, 
we have  $\beta_{\lambda_{\rm eff}}\propto \beta_g$, 
and \eq{intromass} predicts $m^2_{\phi_d}\propto\beta_g$.
Therefore, a dilaton can be parametrically light  if  the dimension of ${\cal O}_g$ is given by $4+\delta$
with $\delta\ll 1$ ({\it i.e.}, $\beta_g\ll 1$) being a controllable small parameter till the end of the RG-flow.
 The holographic implementation of this is the  Goldberger-Wise mechanism \cite{Goldberger:1999uk},
 where the operator ${\cal O}_g$  matches to an almost massless scalar in 5D (protected by a shift symmetry) \cite{GW,Megias:2014iwa}.
 Nevertheless, we will see that  this is not the case at the conformal transition, as the marginal operator ${\cal O}_g$ corresponds to a double-trace operator whose dimension is not protected along the RG-flow.
Having the explicit breaking of conformal invariance arising from an almost marginal operator however
will have as a consequence  that the dilaton is light, although not parametrically light.

We will be   working  with a  simple weakly-coupled AdS$_5$ theory,
with the extra-dimension cut off  by an IR-brane,
 that will contain the basic ingredients to describe the conformal transition.
We will calculate the mass spectrum of resonances  and 
 show that the lightest resonance is the dilaton (the radion of the compact extra-dimension).
We will present a simple analytical formula for the mass of the dilaton that will 
allow to understand its lightness as a function of the change of the tachyon as we move the IR-brane.
This will show that  either at small or large positions of the IR-brane,
the dilaton is always parametrically light. 
In between  these two regions, we will see that the dilaton mass does not have "room" to grow 
and as a consequence the dilaton  is always kept light.

We will compare our results with lattice simulations,  showing good  agreement in the pattern of masses
when the conformal critical point is approached.
Furthermore, we will provide further predictions to be checked  in the future by  lattice simulations.

The 5D model presented  here could also be useful to generate small scales
and   explain, for example, the difference between the electroweak scale and the Planck scale.
Moreover, the presence of  a  light scalar 
can have an important impact in the searches   for  new resonances at the  LHC as
predicted in  composite Higgs models.

There have been previous  approaches using holography 
to understand the conformal transition and the existence of a light dilaton \cite{Kutasov,lightdilaton}.
We find however  that these  studies were not exhaustive nor conclusive.
Our goal is not only  to  provide evidence for a relatively  light dilaton, but also 
to explain the  reasons behind this.

The article is organized as  follows. 
In Section~\ref{ini} we introduce the idea of leaving    the conformal window  by fixed-point merging and 
remark its implications.
In Section~\ref{model5D} we present the five-dimensional model and
 its relation with the large $N_c$ and $N_F$ expansion.
 We also discuss the  tachyon solution and  the stabilization of the radion.
Next  we present the predictions for the resonance mass spectrum,
 presenting an analytical formula for the case when the dilaton is light,
 as well as  discussing  the other scalar and vector resonance masses.
In Section~\ref{comparison}
 we compare the mass spectrum calculated within our model with that obtained from lattice simulations,
and in Section~\ref{hierarchy} we  discuss how  these models could also be  useful for explaining the smallness of the electroweak scale.
Conclusions are given in Section~\ref{conclusions}.
We also present two Appendices. In Appendix~\ref{appA}
we give the coupled system of equations of motion for the scalar and gravitational sectors,
and derive the approximate analytical formula for the dilaton mass.
In Appendix~\ref{appB} we present the 4D effective theory of a tachyon and dilaton
valid when  they are the lightest states.

\section{Conformal transition by fixed-point merging}
\label{ini}

\begin{figure}[t]
\centering
\includegraphics[width=0.45\textwidth]{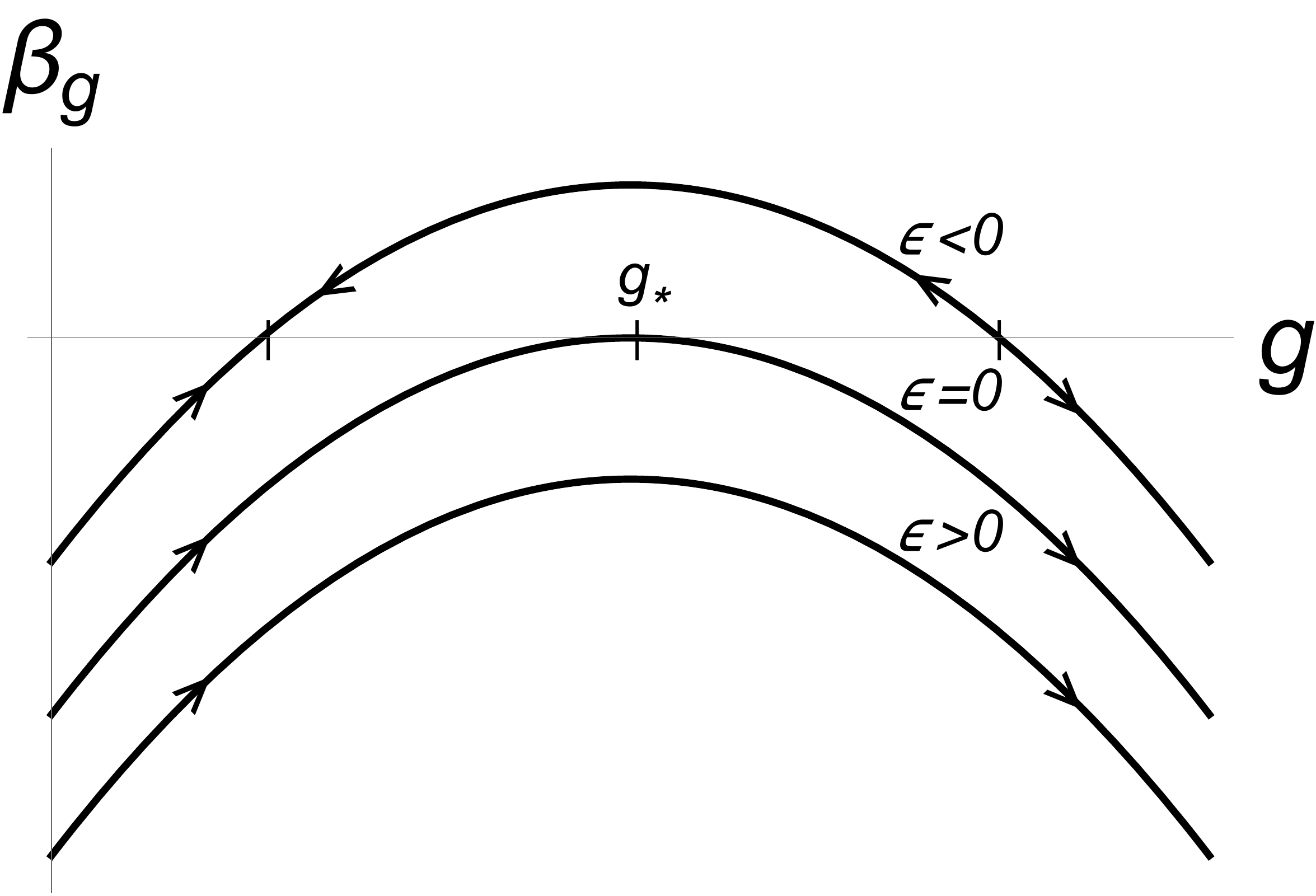}
\caption{\it Beta function of the coupling  $g$ for different values of $\epsilon$. For $\epsilon=0$, the IR and UV fixed points merge at $g_*$.}
\label{merging}
\end{figure}

There are several ways to lose an IR fixed point as we move the parameters of the theory.
Either the fixed point goes to zero, to infinity or it merges  with a UV fixed point.
Following Ref.~\cite{Kaplan:2009kr} we will consider
 conformal transitions  characterized by the third case, the merging of the IR fixed point  with a UV fixed point,
 as depicted in Fig.~\ref{merging}.
In this case, the beta function  can be written as
\be
\beta_g\simeq -\epsilon-(g-g_*)^2\,,
\label{betag}
\ee
where $g$ is a coupling of the theory (not necessarily related to the gauge coupling in gauge theories),
and $\epsilon$ depends on the parameters of the theory, e.g., $N_F$. The IR and UV fixed point are respectively at
\be
g=g_*\mp \sqrt{-\epsilon}\,.
\ee
As we vary  $\epsilon$  from negative to positive values, 
we have   the merging of the IR and UV fixed points at  $\epsilon=0$, while 
for  $\epsilon>0$ the theory abandons conformality, {\it i.e.}, 
the IR fixed point is at   complex coupling.

As argued in Ref.~\cite{Gorbenko:2018ncu}, for $\epsilon$ negative and close to zero,
the operator ${\cal O}_g$ with coupling  $g$
must have dimension 
\be
{\rm Dim}[{\cal O}_g]=4+\frac{d\beta_g}{dg}\simeq4+2\sqrt{-\epsilon}\,,
\label{complexdim}
\ee
and can be considered to be responsible for the RG flow towards the  IR fixed point.
For $\epsilon=0$ we have that   ${\cal O}_g$ becomes marginal,
and develops a complex dimension for  $\epsilon>0$,  signaling the end of conformality.

The above properties of this conformal transition 
have a straightforward holographic  interpretation
using 
the correspondence (or duality) between strongly-coupled CFT$_4$ (in the large $N_c$ and large 'tHooft coupling)
and  weakly-coupled five-dimensional Anti-de-Sitter theories (AdS$_5$) \cite{adscft}.
Operators  in the CFT$_4$ (${\cal O}$) correspond to scalars in the AdS$_5$  ($\Phi$)
where dimensions and masses are related via the  the AdS/CFT relation \cite{adscft}:
\be
{\rm Dim}[{\cal O}]=2+\sqrt{4+M^2_\Phi L^2}\,. 
\label{dic1}
\ee
\eq{dic1} tells us that in order to have a dual of a CFT operator with    complex dimension,
the  AdS$_5$ must have a scalar  slightly  below  the BF-bound,   $M^2_\Phi=-(4+\epsilon)/L^2$. 
\eq{dic1} also tells us that 
this operator  of complex dimension  (${\cal O}_*$)  has, in the limit  $\epsilon\to 0$,  dimension $2$ instead of $4$. 
Therefore the natural identification for 
the ${\cal O}_g$ operator discussed above is ${\cal O}_g=|{\cal O}_*|^2$,  since in the large $N_c$ this implies
 ${\rm Dim}[{\cal O}_g]=2{\rm Dim}[{\cal O}_*]$ and that gives us \eq{complexdim}.
In other words, ${\cal O}_g$  must be a   double-trace operator.\footnote{It has  been proven in Ref.~\cite{Pomoni:2008de,Gorbenko:2018ncu} that this is always the case  for  theories in  the large-$N_c$ limit.}

The existence of  ${\cal O}_*$ in the conformal transition is only an implication from large-$N_c$ theories, and could
be not true in general. In QCD, as argued in Ref.~\cite{Kaplan:2009kr},
 ${\cal O}_*$  is  expected  to be the $q\bar q$ operator whose dimension will  go from $\sim 3$ 
 when entering the conformal window (at the Banks-Zaks fixed point) 
  to 2  when  leaving it at the other side when it becomes complex.

When the theory is close but outside  the conformal window ({\it i.e.} $0<\epsilon\ll 1$), one can calculate the RG flow ``time" required to  cross the region where $g\sim g_*$  and $|\beta_g|\ll1$.
This gives us the IR-scale $\Lambda_{\rm IR}$ at which the
theory is expected to confine as $g$ becomes large.
 From \eq{betag} one gets
\be
\Lambda_{\rm IR}\sim e^{-\pi/\sqrt{\epsilon}}\Lambda_{\rm UV}\,,
\label{lambdarelation}
\ee
where $\Lambda_{\rm UV}$ is roughly the scale at which $g\lesssim g_*$.
\eq{lambdarelation} is usually referred as {\em walking} or {\em Miransky} scaling.

\section{A  five-dimensional model for the conformal transition}
\label{model5D}

We  want to study the conformal transition described above using  holography.
For this reason we  will consider  the simplest but at the same time most generic  five-dimensional model containing
 the basic ingredients needed to describe the conformal transition via  fixed-point merging.
Our purpose  is to  generically understand   the mass spectrum at the 
 conformal transition and the presence or not of light scalars.

Let us recapitulate the basic ingredients of the theory in the 4D side.
This is a   strongly-coupled deformed CFT with a scalar operator,  
$q^i_L\bar q^j_R$ ($i,j=1,..., N_F$) for concreteness,
whose dimension is  $2+\sqrt{-\epsilon}$ with $0<\epsilon\ll 1$.
This means that the scalar $q^i_L\bar q^j_R$  gets a vacuum expectation value (VEV),
signaling the loss of conformality.
The  global symmetry of this theory is  $SU(N_F)_L\otimes SU(N_F)_R\otimes U(1)_B$
that is  broken by the  VEV of the  scalar $\langle q^i_L\bar q^j_R\rangle\propto \I$ down to the diagonal subgroup 
$SU(N_F)_L\otimes SU(N_F)_R\to SU(N_F)_V$.\footnote{The U(1)$_A$  is anomalous and will not be considered here.}

The corresponding  holographic model will  consist in a    $SU(N_F)_L\otimes SU(N_F)_R\otimes U(1)_B$ gauge theory in 5D with a complex scalar $\Phi$ 
transforming as a  $({\bf N_F}$,${\bf\overline  N_F})_{\bf 0}$.
This scalar plays the role of the  $q\bar q$  operator whose VEV is responsible for the
breaking of the conformal and gauge symmetry, and therefore its mass will be related to the dimension of the $q\bar q$  operator.
We also impose parity,   defined as the interchange $L \leftrightarrow  R$. 
 The action is given by 
\begin{equation}
S_5=\int d^4x\int dz\, \sqrt{g}\,  M_5 \left[\frac{1}{\kappa^2}\left({\cal R}+\Lambda_5\right)+{\cal L}_5\right]\, ,
\label{s5}
\end{equation}
where, up to dimension-four operators,\footnote{Following the Effective Field Theory (EFT) approach,  higher-dimensional operators are supposed to be suppressed by the cutoff scale  of the model ($\Lambda_{\rm cutoff}$) estimated to be  the scale at  which the 5D theory  becomes strongly coupled ({\it i.e.}, when loops are as important as tree-level contributions),
that is    $\Lambda_{\rm cutoff} \sim 24\pi^3 M_5$  \cite{Ponton:2012bi} -- see also Section~\ref{largenssec}.} the most general Lagrangian is given by
\begin{equation}
{\cal L}_5 = 
-\frac{1}{4}\Tr \left[ L_{MN}L^{MN}+R_{MN}R^{MN} \right]
-\frac{1}{4} B_{MN}B^{MN}
 +\frac{1}{2} \Tr |D_M\Phi|^2
-V_\Phi(\Phi)\,,
\label{la5}
\end{equation}
with $L_{MN}$, $R_{MN}$ and $B_{MN}$ being the field-strength of the $SU(N_F)_L$,  $SU(N_F)_R$
and $U(1)_B$ gauge bosons respectively,  and the indices  run  over the five dimensions, $M=\{\mu,5\}$.
We parametrize the fields
as $\Phi=\Phi_s+T_a\Phi_a$
 with  Tr$[T_aT_b]=\delta_{ab}$.
 The fields $\Phi_s$ and $\Phi_a$ will respectively transform as singlet and adjoint under the $SU(N_F)_V$.
The  covariant derivative is defined as
\begin{equation}
D_M\Phi=\partial_M \Phi+ig_5L_M\Phi-ig_5\Phi R_M\, ,
\end{equation}
and the potential is given by\footnote{We notice that one can absorb  one coupling into $M_5$, as we will do later.}
\be
V_\Phi(\Phi)=\frac{1}{2}M^2_\Phi\Tr|\Phi|^2+\frac{1}{4}\lambda_1 \Tr|\Phi|^4+\frac{1}{4}\lambda_2 (\Tr|\Phi|^2)^2\, .
\label{pot5D}
\ee
The 5D metric in conformal coordinates is defined as
\begin{equation}
ds^2=a(z)^2\big(\eta_{\mu\nu}dx^\mu dx^\nu-dz^2\big)\, ,
\label{metric}
\end{equation}
where   $\eta_{\mu\nu}=\diag(1, -1, -1, -1)$ and 
$a(z)$ is the warp factor. Before the scalar $\Phi$ turns on, the presence of $\Lambda_5$ leads to  an     AdS$_5$ geometry: 
\begin{equation}
a(z)=\frac{L}{z}\,,
 \label{ads}
\end{equation}
where  $L^2=12/\Lambda_5$ is the squared AdS curvature radius.

As explained above, our important ingredient  here  is to consider that the conformal breaking arises
when  Dim[$q\bar q$]   becomes imaginary. 
In AdS this corresponds from \eq{dic1} to take the
  5D mass of $\Phi$     below the BF bound.
 For this purpose, we will consider
\be
M^2_\Phi=-\frac{4+\epsilon}{L^2}\,.
\label{mass5D}
\ee
When  $\epsilon>0$ the  mass of $\Phi$ is below the BF bound and   $\Phi$  turns on in the 5D bulk,
breaking the conformal and chiral symmetry $SU(N_F)_L\otimes SU(N_F)_R\to SU(N_F)_V$.
$\Phi$ will grow as $\sim z^2$, as expected from a dimension-two perturbation in the dual 4D theory.
 When the energy-momentum tensor induced by the nonzero $\Phi$ profile  gets of order of the inverse 5D Newton constant, $\kappa^2$, the backreaction on the metric will be important, starting to depart then from AdS,
and signaling the breaking of the conformal symmetry.
This simple model, however, does not lead to a mass gap for all bulk fields,  as the extra dimension
is not ending at any $z$. In fact, as we will see, the tachyon $\Phi$ would stabilize at the minimum
of the potential \eq{pot5D} and  the metric would become again AdS. 
As we know that in strongly-coupled models outside the conformal window, like QCD,  all resonances are heavy, we need
to implement the same in  our holographic version.  The simplest way is to
cut off the 5D space by an  IR-brane at some point   $z=\zir$ to be  determined dynamically.

The presence of  the IR-brane add extra parameters to the theory as
$\Phi$ might also have a potential on the IR-boundary. 
We will limit ourselves to up to quadratic terms in $\Phi$:
\begin{equation}
{\cal L}_{\rm IR} = -a^4 \tilde V_b(\Phi)\big|_{\zir} ,\qquad \tilde V_{b}(\Phi)=\frac{\Lambda_4}{\kappa^2}+\frac{1}{2}m^2_b\Tr|\Phi|^2\,, 
\label{pot4D}
\end{equation}
and study their  impact on the properties of the model. 
We could also  add  to \eq{pot4D} quartic terms but these are not expected to 
 change significantly our predictions, since
  they are   suppressed by $1/(M_5L)$   with respect to the bulk terms.\footnote{Even though the same is true for the two terms in \eq{pot4D},    these are respectively quartically and quadratically  UV sensitive so they can be sizable.}
  As it is usual in AdS/CFT, we will be regularizing the UV-divergencies by placing a UV-boundary
at $z=\zuv$ and taking the limit $\zuv\to 0$ at the end of the calculation of physical quantities.

In this 5D model the two phases are  determined, as in the strongly-coupled model described in Sec.~\ref{ini},
 by the $\epsilon$ parameter:
\begin{itemize}
\item For $\epsilon<0$, we have $\Phi=0$ and $\zir=\infty$: AdS$_5$  (CFT$_4$) phase.
\item For $\epsilon>0$,  we have $\Phi\not=0$ and $\zir\not= \infty$: non-AdS$_5$ (non-CFT$_4$) phase.   
\end{itemize}

\subsection{The large $N_c$ and $N_F$ power counting}
\label{largenssec}

By the  AdS/CFT correspondence, 
the 5D scalar and gauge bosons are associated  to the meson operators $\bar q q$ and $\bar q\gamma_\mu q$ respectively.
Therefore  5D couplings from single-trace operators  must scale in this sector as $1/N_c$.
We take
 \be
 \frac{1}{M_5L}\sim \frac{16\pi^2}{N_c}\ ,\ \ \ 
 \lambda_1\sim g_5\sim N_c^0\,.
 \label{largens}
 \ee
On the other hand,   double-trace operators are suppressed  with respect to single-trace ones:
\be
\lambda_2\sim \frac{1}{N_c}\,.
\ee
For this reason these latter terms were neglected  in previous holographic approaches to QCD \cite{Erlich:2005qh,DaRold:2005mxj}. Nevertheless,  the parameter  $\lambda_2$ is accompanied by  a factor  $N_F$, as we will see explicitly below (e.g.  \eq{l2}), and then its effect is not suppressed  for large values of $N_F$.
Therefore it is important to keep double-trace operators in \eq{la5}
when comparing our results to  strongly-coupled  theories in the large $N_c$ and $N_F$ limit.
In particular, $\lambda_2$ will be responsible for  generating a mass splitting in the scalar sector between the singlet ($\Phi_s$)
and the adjoint states ($\Phi_a$), as observed in lattice results with large $N_F$ \cite{Aoki:2016wnc,Brower:2015owo}.

It is important to remark that we cannot consider the strict limit $N_F\sim N_c$ in our 5D model.
In this  limit  loops of vector or scalar resonances   contribute as  $\frac{1}{16\pi^2}\frac{N_F}{M_5L}\sim \frac{N_F}{N_c}\sim 1$,
meaning that we cannot perform a perturbative expansion in the 5D theory.
Therefore calculations   will be only reliable if  we take the large $N_F$ and large $N_c$ limit ($M_5L\to \infty$)
keeping  ${N_F}\ll {16\pi^2}{M_5L}$.
Basically, the only difference here with respect to previous models for holographic QCD
is the non-negligible presence of  double-trace operators.

On the other hand,  if we limit ourselves to the
flavor-singlet sector of the theory  (assuming for example that the other sectors are heavier),
this can be treated perturbatively even in the strict Veneziano limit $N_F\sim N_c\to \infty$.
It has been argued in Ref.~\cite{Gadde:2009dj} that  
 there should be a dual purely closed string description of the flavor-singlet sector of the gauge theory.

From the AdS/CFT dictionary, we are also able to relate the gravitational sector of the 5D theory with the glueball sector
of the 4D CFT,  and derive the scaling of the  5D Newton constant with the number of colors:
$\kappa^2/(M_5L^3)\sim 16\pi^2/N_c^2$ that using \eq{largens} implies
\be
\frac{\kappa^2}{L^2}\sim \frac {1}{N_c}\,.
\ee
From the above we can estimate  the mixing between the flavor-singlet meson sector and the glueball sector
(dual  respectively to the scalar and gravitational sectors  in 5D)  to go as  
\be
\hat\kappa^2\equiv \frac{ \kappa^2N_F}{L^2} \sim \frac{N_F}{N_c}\,,
 \ee
that  becomes order one for $N_F\sim N_c$.
 Therefore, contrary to previous holographic models,  
  the impact of the gravitational sector in the singlet scalar sector cannot be  neglected in this case.
 
\subsection{The tachyon solution}
\label{tasol}

The non-zero profile for $\Phi$ will be taken to be along the  $\phi=|\Phi_s|$ direction, whose 5D Lagrangian 
is given by
\begin{equation}
{\cal L}_{\phi} = 
N_F\left[\frac{1}{2} (\partial_M\phi)^2-V(\phi)\right]+g^2_5\phi^2\Tr A_{M}^2\,,
\label{la5s}
\end{equation}
where
\be
V(\phi)=\frac{1}{2}M^2_\Phi \phi^2+\frac{1}{4}\lambda \phi^4\ ,\ \ \   \ \      \lambda\equiv\lambda_1+{N_F}\lambda_2\,,
\label{l2}
\ee
being $A_{M}=(L_M-R_M)/\sqrt{2}$   the axial-vector gauge bosons that will get  masses
from their coupling to $\phi$. 
The IR-brane potential can be written as
\be
{\cal L}_{\rm IR} = -a^4N_F  V_b(\phi)\big|_{\zir} ,\qquad 
V_b(\phi)=\frac{\Lambda_4}{\hat \kappa^2L^2}+\frac{1}{2}m^2_b\phi^2\,.
\ee
Notice that the presence of a factor $N_F$ in front the 
Lagrangian means that the couplings of $\phi$ are   suppressed by an extra $1/N_F$ with respect
to those in the non-singlet sector, as
 expected  in strongly-coupled theories in the large $N_c\sim N_F$ limit.
The equation of motion (EOM) for $\phi$ from \eq{la5s} must be solved  
 including the metric back-reaction that via the   Einstein equations (see Appendix~\ref{appA})
 determines  the warp factor:
\be
-\frac{\dot a}{a^2}=\sqrt{\frac{1}{L^2}+\frac{\hat\kappa^2L^2}{12}\Big(\frac{\dot{\phi}^2}{2a^2}-V(\phi)\Big)}\,,
\label{wf}
\ee
where from now on we  will be using  the dot notation: $\dot \phi\equiv\partial_z \phi$.
It is important to notice  that   by the field redefinition   $\phi\to \phi/\sqrt{|\lambda|}$
we could factorize $|\lambda|$ in front of the first term of \eq{la5s} and  make the
EOM that determines the solution for  $\phi$ independent of $|\lambda|$ (only sensitive to its sign).  
This redefinition introduces $|\lambda|$ in  the interactions of $\phi$ with the  gauge  and  gravitational  fields
 (second term of 
\eq{la5s} and \eq{wf} respectively).  Nevertheless, this 
 can be reabsorbed respectively in  $g^2_5$ and $\hat\kappa^2$, making 
the solutions and full  mass spectrum  of the model insensitive to $|\lambda|$. 
Therefore, with no loss of generality, we will consider $\lambda=\pm 1$.

We are interested to study the model close to the  conformal transition.
Therefore  we will work in the limit  $\epsilon\to 0$. 
The solution for $\phi$ then only depends on  $\zir$, $\hat \kappa^2$ and $m^2_b$ (and the sign of $\lambda$).
At the UV-boundary we will impose $\phi=0$, otherwise
we would be breaking the chiral symmetry  from UV-physics 
(as adding an explicit  mass term to the quarks in the dual theory).\footnote{Imposing a different boundary condition, such as $z\dot\phi |_{\zuv}\propto \phi |_{\zuv}$, would lead to the same predictions in the limit  $\epsilon\to 0$ ($\zuv\to 0$).}
On the other hand, at the IR-brane  we must impose the boundary condition determined by the model:
\be 
\left. \left(\frac{M_5}{a} \dot\phi +V'_b\right)\right|_{\zir}=0\,,
\label{bc}
\ee
where we defined  $V'_b\equiv \partial_\phi V_b$.
For the metric we must impose the junction condition \cite{Csaki:2000zn}:
\be
\left.  \left(-\frac{6 M_5}{\hat \kappa^2L^2}\frac{\dot a}{a^2}+ V_b\right)\right|_{\zir}=0\,.
\label{junction}
\ee

\subsubsection{Region $\hat m_b^2> -2$} 
We will start looking for solutions of the tachyon for $\hat m_b^2> -2$, 
where  we define
\be
 \hat m_b^2\equiv \frac{m_b^2L}{M_5}\,.
 \ee
In this case non-trivial solutions from \eq{la5s} fulfilling  \eq{bc} are only  found  if the IR-brane is beyond some critical value, $\zir>\zirc$.
It is easy to find $\zirc$, as this corresponds to the critical value at which we pass  from having all Kaluza-Klein (KK) states of $\phi$ with positive squared masses    to having  4D tachyons  in the theory.
Therefore at  $\zir=\zirc$ there must be  a  4D massless mode, $\phi_t(x)$.
The wave-function of this massless mode must satisfy the linearized bulk EOM with $p^2=0$. 
We obtain \cite{Kutasov}
\be
\phi(x,z)=\frac{\phi_t(x)}{N}\, z^2 \sin\left(\sqrt{\epsilon} \ln\frac{z}{\zuv}\right)\,,
\label{eqtachyon}
\ee
with $N$  a normalization constant, and where the IR-boundary condition  \eq{bc} at $\zir=\zirc$ leads to
\be
\tan\left(\sqrt{\epsilon} \ln\frac{\zirc}{\zuv}\right)=-\frac{\sqrt{\epsilon}}{2+\hat m^2_b}\ \ \  \ \Rightarrow\ \  \  \ \sqrt{\epsilon} \ln\frac{\zirc}{\zuv}\simeq n\pi\ ,\ \  n=1,2,...\,.
\label{zirc}
\ee
Notice that to have non-trivial solutions, the limit $\epsilon\to 0$ must be taken with $\zuv\to 0$, such that the angle 
in \eq{eqtachyon} is kept fixed.
The presence of $n$ solutions in \eq{zirc} is 
a  well-known feature of these configurations,  and it is  associated to the existence of   Efimov states.
We will be considering $n=1$, that will give us the global minimum, being the other possibilities just local minima. 
\eq{zirc} reproduces \eq{lambdarelation} for $\Lambda_{\rm IR}\sim 1/\zirc$ and
$\Lambda_{\rm UV}\sim 1/\zuv$. 
The origin  of the logarithm in \eq{eqtachyon}, that  will play an important role,
 can be more easily understood by looking at the strongly-coupled dual theory;
this has   an explicit breaking of the conformal symmetry due to the double-trace marginal operator 
${\cal O}_g=|{\cal O}_*|^2$ that leads to a  log-running of the couplings \cite{Witten:2001ua}.

\begin{figure}[t]
\centering
\includegraphics[width=.93\textwidth]{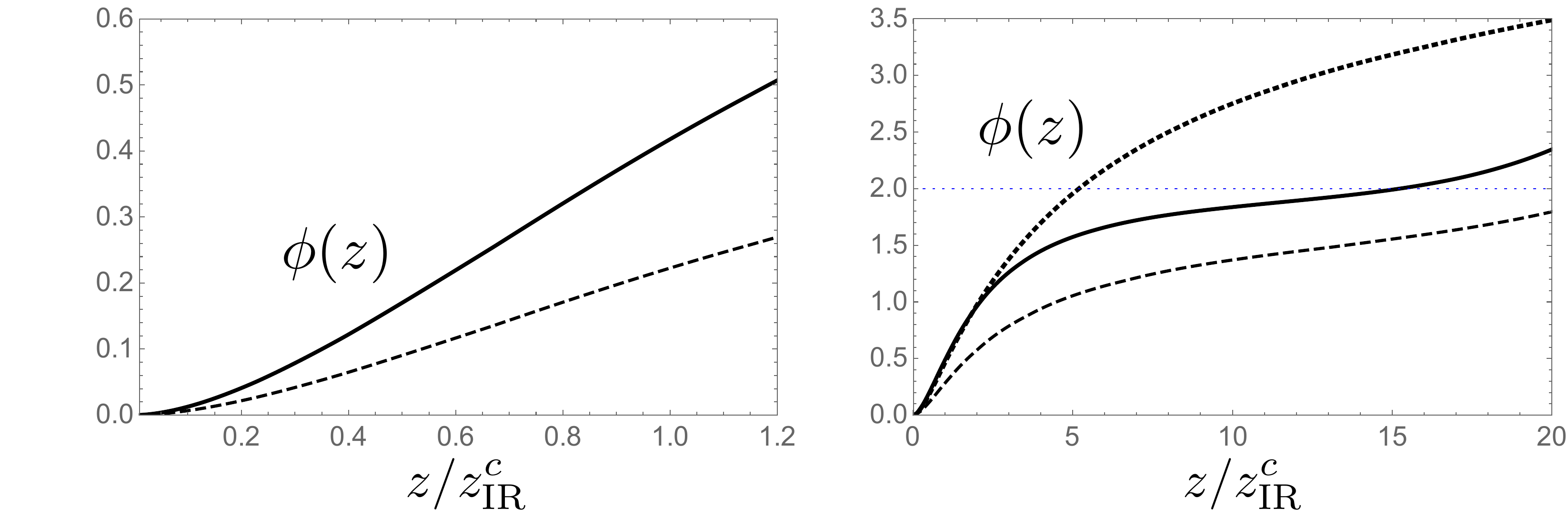}
\caption{\it 5D tachyon solutions (in units of $1/L$) 
for $\hat m^2_b=-1$. 
LEFT:    limit  I with $\zir= 1.2\,\zirc$. RIGHT:   limit  II with $\zir= 20\, \zirc$. 
We have taken  $\lambda=1$,  and  $\hat\kappa^2=1\ (4)$ for the solid (dashed) line,
and $\lambda=-1$, $\hat\kappa^2=4$ for the dotted line.}
\label{tachyonprofile}
\end{figure}

Depending on the position of the IR-brane with respect to $\zirc$, we can distinguish two limiting cases 
 that will help to understand the physics of the model. These are  illustrated in Fig.~\ref{tachyonprofile} and corresponds to
\begin{enumerate}[I)]
\item $\zir\approx\zirc$:  In this case  $\phi L\ll 1$  for all $z$, meaning 
that  the scale of confinement $\sim 1/\zir$
is larger than the  scale of chiral breaking that is of order $\sim L\phi(\zir)/\zir$.
\item  $\zir\gg  \zirc$: In this case $\phi L$ reaches  $O(1)$ values at some $z=z_{\chi}\ll \zir$, 
and then the scale of chiral breaking  $\sim 1/z_{\chi}$  is larger than the scale of confinement $\sim1/\zir$.
\end{enumerate}
These two cases should be considered as formal limits, since in most of the parameter space of the model
 we will find that the IR-brane  sits at  $\zir\sim $ few $\times\ \zirc$,
{\it i.e.}, between   limits I and II, 
implying that naturally  the scale of chiral breaking is similar to the scale of confinement.

Let us start considering the limit  I, where  $\zir$ is assumed to be just slightly above $\zirc$.
In this case   the 4D  mode $\phi_t(x)$  gets  a small  negative mass-squared, 
becoming  a  4D tachyon.
We  find this mass is given by
\be
m^2_t\simeq -\frac{\beta}{\zir^2} \ln\frac{\zir}{\zirc}\ ,\  \ {\rm where}\ \ \
\beta=\frac{4 (\hat m_b^2+2)^2}{\hat m_b^4+6\hat m_b^2+10}\,.
\label{tachy}
\ee
 \eq{tachy}  is only valid for $|m^2_t|\ll 1/\zir^2$ that, obviously,
requires a tuning in the  parameter space:
either  $\ln{\zir}/{\zirc}\ll 1$ (that we will see later  cannot be achieved by the radion minimization)
or  $\beta\ll 1$ that requires  $\hat m^2_b\to -2$.
To find a stable configuration this 4D tachyon must have a positive quartic self-interaction, $\lambda_t>0$,
and this can arise either from $\lambda$  or the feedback from gravity.
We find
\be
\lambda_t=\lambda\,  c_\lambda(\hat m^2_b)+\hat\kappa^2c_\kappa(\hat m^2_b)\,,
\label{effla}
\ee
 where   $c_{\lambda,\kappa}$  are  smooth and positive functions of $\hat m^2_b$ as derived in Appendix~\ref{appQuartic}.
The 4D tachyon VEV is then given by
\be
\langle \phi_t\rangle= \frac{1}{\zir}\sqrt{\frac{\beta}{\lambda_t} \ln\frac{\zir}{\zirc}}\,.
\label{phit}
\ee
Therefore, in the limit I   the $z$-profile of $\phi$  is given by \eq{eqtachyon} with \eq{phit} and $N=L\zir\epsilon/\beta$,
as we follow the normalization of $\phi_t$ of  Appendix~\ref{appQuartic}.
This solution is shown in the left plot of Fig.~\ref{tachyonprofile} for $\lambda=1$ and $\hat\kappa^2=1,4$.
In the limit $\hat\kappa^2\gg 1$, we see from  Eqs.~(\ref{eqtachyon})--(\ref{phit}) that $\hat \kappa^2\phi^2L^2$
stays constant and small. This means that   the metric remains always  close to AdS$_5$.

Let us now move to  the limiting case  II. First, let us neglect  the feedback from the metric ($\hat\kappa^2\ll 1$).
As  the IR-brane is now  placed  far away from $\zirc$,
  the tachyon profile  grows  $\propto z^2$ till   the quartic term of the potential  becomes relevant.
Solutions only exist  if  $\lambda>0$,  such that $\phi(z)$ settles at 
 the minimum of the 5D potential $V(\phi)$ where it takes the constant value $\phi(z)\simeq \sqrt{{M^2_\Phi}/{\lambda}}=2/(L\lambda)$ (see right plot of Fig.~\ref{tachyonprofile}).\footnote{\label{footy} To satisfy the boundary condition at the IR-brane, $\phi$ must depart from $2/(L\lambda)$ when approaching the IR-boundary,  as can be appreciated in Fig.~\ref{tachyonprofile}.}
 This means, in the dual interpretation,
that the CFT  flows at around $1/\zirc$ towards   another CFT  in which
the global symmetry has been reduced to   $U(N_F)_V$ with  $\Phi_s$  and $\Phi_a$ respectively
transforming  in the singlet and  adjoint representation.
In this new CFT, scale invariance is broken at a much lower scale $1/\zir$.
Let us now consider the feedback of the metric. For large $\hat\kappa^2$
the gravitational feedback  becomes important before $\phi$ reaches  the $V(\phi)$   minimum, 
making $\phi$ to   enter  into a "slow-roll" condition (see Appendix~\ref{appA} for details)  delaying the position $z$ at which
$\phi$ gets its maximum $\sim 2/(L\lambda)$.
In this case $\lambda<0$   is also possible as the slow-roll condition keeps $\phi(z)$  slowly growing till reaching the IR-brane
(see   right plot of Fig.~\ref{tachyonprofile}).
The metric evolves from AdS$_5$ at  $z\approx\zirc$ to  another approximately AdS$_5$ space at  $z\gg \zirc$.

\subsubsection{Region $\hat m_b^2< -2$} 
\label{minus2region}
In this region we have that $2+\hat m^2_b$ is negative,  and from the left-hand side of \eq{zirc}, the smallest $\zirc$ is determined   by 
\be
\ln\frac{\zirc}{\zuv}\simeq-\frac{1}{2+\hat m^2_b}\,,
\label{zirc-2}
\ee
that does not depend on  $\epsilon$. This means that non-trivial solutions for $\phi$ exist 
 even if $\epsilon<0$.
 These solutions however are supported by the IR-brane and  for $\zir\to\infty$ we have $\phi\to 0$.
Therefore as soon as the IR-brane is not stabilized for $\epsilon<0$ ({\it i.e.}, $\zir\to\infty$), we can also consider this region of the parameter space for studying the conformal transition.

In this case the solution for $\phi$,  as we vary $\zir$, behaves in the following way.
For $\zirc<\zir<\zircp$, where   $\zircp$ is  determined by   $ \ln({\zircp}/{\zuv})\sim \pi/\sqrt{\epsilon}$,
  we find that  $\phi$ takes a nonzero  value with a profile localized
towards the IR-brane, $\phi\sim (z/\zir)^2$, as we said.
 The origin of this nonzero profile is  that the IR-brane mass $\hat m^2_b$, and not the 5D mass, is exceedingly negative.
 $\phi(\zir)$ is mostly constant  in this region and it does not help to stabilize the IR-brane.
On the other hand, for $\zir>\zircp$,    the profile of  $\phi$ grows to   become similar to the limit II  discussed before (see right-hand side of Fig.~\ref{tachyonprofile}), indicating  that   $\phi$ behaves as a  genuine 5D tachyon. This latter behavior only occurs
if the 5D mass is below the BF bound and can lead to a stable IR-brane.

\subsection{Radion/Dilaton stabilization}
\label{sectiond}

Since the position of the IR-brane $\zir$  is associated to a  dynamical field, the radion (not necessary a mass eigenstate),  its value must be determined dynamically. 
The extremization   condition for $\zir$ 
is exactly  the junction condition \eq{junction} after putting on-shell all other fields.
This can be written, using \eq{wf} and \eq{bc}, as
\be
\left. \left(\frac{6M_5}{\hat\kappa^2L^2}  \sqrt{\frac{1}{L^2}+\frac{\hat\kappa^2L^2}{12}\left(\frac{V_b^{'\, 2}}{2M_5^2}-V(\phi)\right)}\,+\,V_b \right)\right|_{\zir}=0\,.
\label{minD}
\ee
 For our particular case, this reduces to a quadratic equation for $\phi(\zir)$:
\be
\frac{4}{L^2}\,\big(\delta\hat\Lambda - \frac{\delta\hat\Lambda^2}{12}\big)+\hat\kappa^2L^2\left[\frac{1}{2}\bar m^2 \phi^2(\zir)-\frac{\bar \lambda}{4} \phi^4(\zir)\right]=0\,,
\label{minD2}
\ee
where have  introduced 
$
\bar m^2\equiv[(\hat m_b^2+2)^2-2\delta\hat\Lambda \,\hat m_b^2/3]/L^2
$,
$\bar\lambda\equiv\lambda+\hat\kappa^2\hat m_b^4 /3$,
and  
\be
\delta\hat\Lambda \equiv \frac{L}{M_5}\, \Lambda_4+6\,,
\ee
 that is a  measure of the detuning of the IR-brane tension away from the AdS$_5$ value.
Their values are bounded to be in the region
\be
0\leq \delta\hat\Lambda\leq 6\,.
\label{bounds}
\ee
The lower bound     arises  from demanding that  for $\epsilon<0$,
the IR-brane is driven to $\zir\to \infty$, such that the theory is in the AdS$_5$ (CFT$_4$) phase.
From   Appendix \ref{appB}, in particular \eq{lambd},
 we see that $\delta\hat\Lambda$ is related to the self-coupling of the dilaton 
and  $\delta\hat\Lambda\geq 0$ comes from requiring a positive dilaton self-coupling.
On the other hand, the upper limit in \eq{bounds} is a more basic (geometrical) requirement: to possibly solve the junction condition even for dynamical solutions that start away from the minimum. If $\delta\hat\Lambda>6$,
 the IR-brane tension $\Lambda_4$ is positive and it is easy to see that there would be no solutions where the IR-brane acts as an IR boundary ({\em i.e.},  a cutoff   of the AdS$_5$ space at  $z=\zir$).
Therefore these regions must be discarded.
\begin{figure}[t]
\centering
\hskip0cm
\includegraphics[width=.94\textwidth]{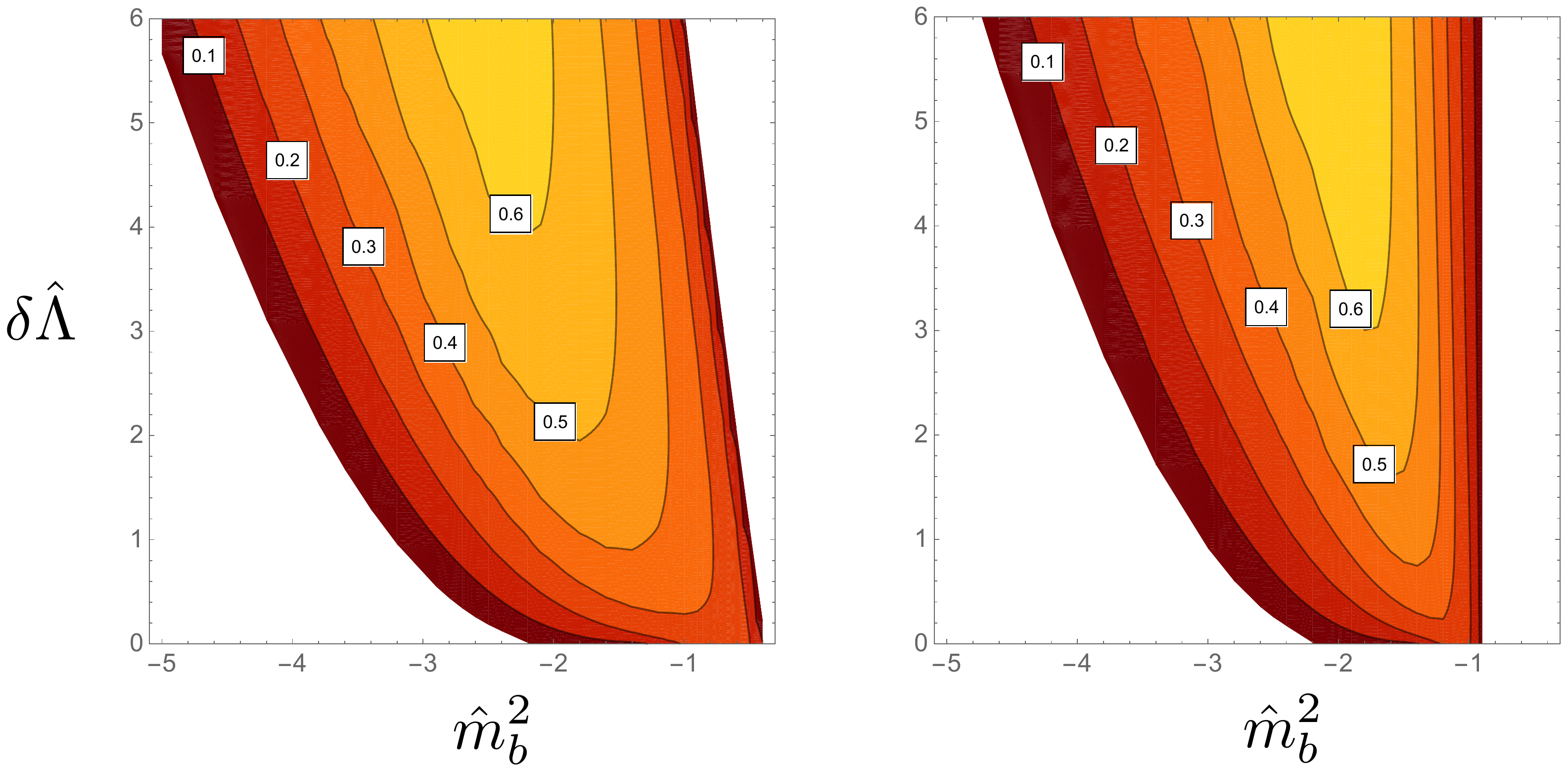} \ \
\caption{\it 
Region of the parameter space
that leads to a stable IR-brane for $\hat\kappa^2=4$ and $\lambda=1$ (left) and $\lambda=-1$ (right). 
We also provide  the value of the lightest scalar mass, $m_{S_1}/m_\rho$.}
\label{islands}
\end{figure}

By playing with the parameters of the model,  $\lambda=\pm 1$, $\hat\kappa^2$, $\hat m_b^2$  and $\delta\hat\Lambda$,   we can find regions where $\zir$ is stabilized thanks to the presence of  the 5D tachyon.
These are  shown in Fig.~\ref{islands} in the plane $\hat m_b^2-\delta\hat\Lambda$ 
for $\hat\kappa^2=4$ and $\lambda=1$ (left plot), and $\lambda=-1$ (right plot).
These  regions are bounded from the left and the right at which, as it will be discussed later, 
the radion is massless.
At the left boundary one obtains the lowest value of (a stabilized) $\zir$.
For  $\hat m^2_b>-2$,  this lowest value of the IR-brane position is achieved when $\delta\hat\Lambda=0$;
for   $\hat m^2_b\to -2$ we obtain the smallest $\zir/\zirc$ that is given by  $\zir=e^{1/2}\zirc$,
as can be analytically found  by looking at the 4D effective theory  (see Appendix~\ref{appB}).
On the other hand, as we get close to the  boundary  on the right  of  the  regions in Fig.~\ref{islands}, 
we  have $\zir\to \infty$  (limit II). 
In most of the  colored regions however we have that  $\zir\sim z_{\chi}$. 
 In other words,  the model naturally predicts the scale of chiral symmetry breaking to be around the scale of confinement.

\begin{figure}[t]
\centering
\includegraphics[width=0.6\textwidth]{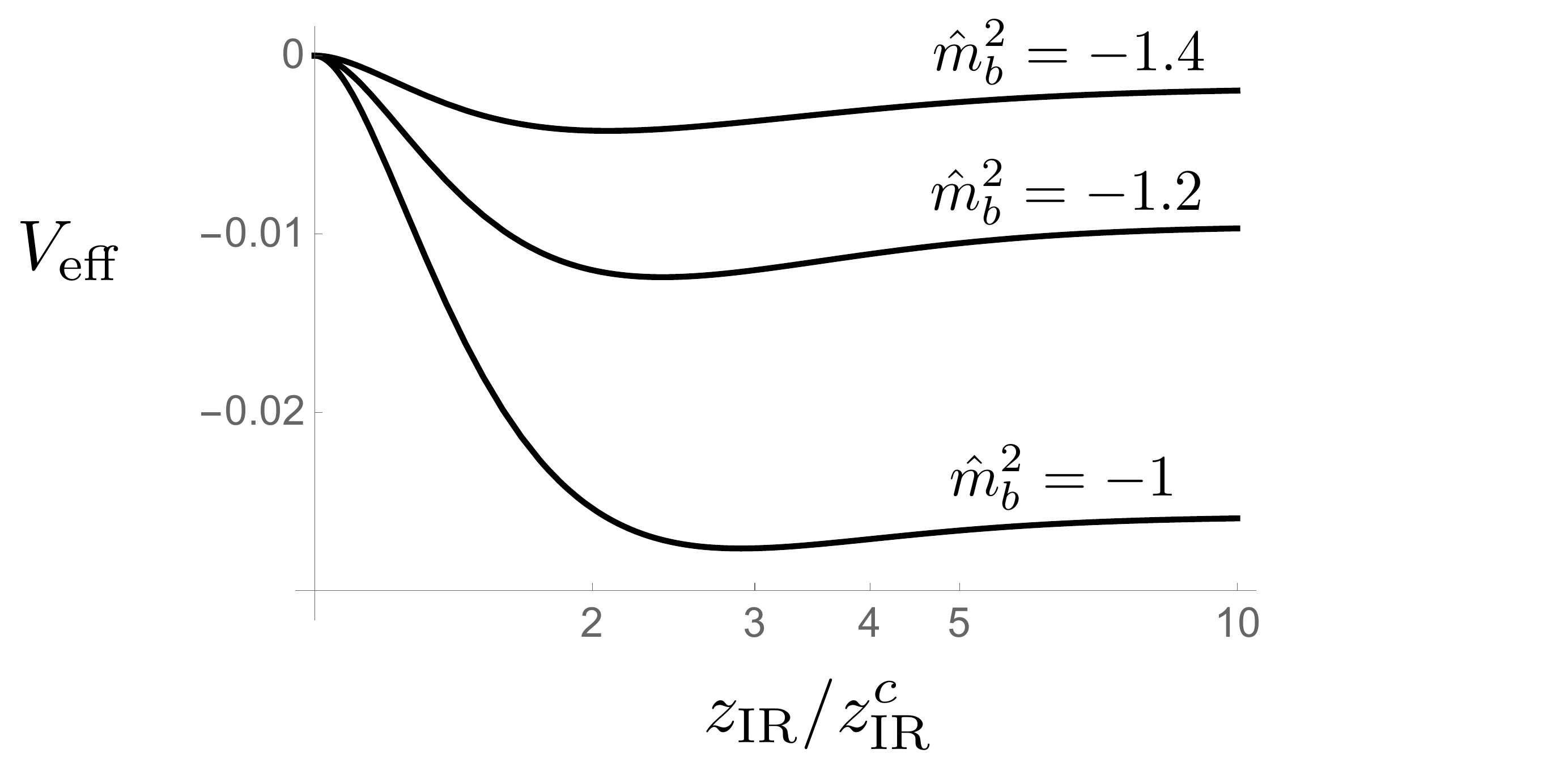}
\caption{\it 
Dilaton effective potential for $\lambda=1$, $\hat\kappa^2=0$,  $\delta\hat\Lambda=0$ and different values of $\hat m_b^2$.}
\label{effpotdilaton}
\end{figure}

If the radion is the lightest 4D mode in the theory, we can use  \eq{minD}
to obtain its effective potential.
The radion corresponds in  the dual 4D CFT to  the dilaton, $\phi_d$,
whose VEV determines the scales of the model.
For this reason  $\phi_d$ at the minimum is  related with the warp factor evaluated at $z=\zir$.
Nevertheless, outside  the minimum  \eq{minD}
 the relation of  $\zir$  with $\phi_d$  is  a more complicated function, $\zir=f(\phi_d)$,
especially in the basis where $\phi_d$ is canonically normalized.
Going off-shell requires not  equating  the LHS  of   \eq{minD}  to zero, 
and identifying this  with the  first derivative of the dilaton effective potential:
\be
\frac{dV_{\rm eff}(\phi_d)}{d\phi_d}=
\left.
 \frac{\phi_d^3}{n(\phi_d)}\left(
\sqrt{1+\frac{\hat\kappa^2L^4}{12}\left(\frac{V_b^{'\, 2}}{2M^2_5}-V(\phi)\right)} \,
+ \,\frac{\hat \kappa^2L^3}{6M_5}\, V_b\right)\right|_{\zir=f(\phi_d)}\,,
\label{minV}
\ee
where   $n(\phi_d)>0$ is in  general a complicated  function of $\phi_d$ (that cannot be zero, otherwise we will have an extra minimum beyond  \eq{minD}) that we do not need to specify here. 
By integrating \eq{minV} over $\phi_d$,  one  can obtain the dilaton effective potential  $V_{\rm eff}(\phi_d)$.
For the simple case  in which  the backreaction is neglected
and the space is just AdS$_5$,   we have  $\zir\propto 1/\phi_d$
and  $n(\phi_d)$ is  just a constant.
 For this case we show the effective potential  (up to an overall constant)   in Fig.~\ref{effpotdilaton}.
We can see that the potential has a minimum at $\zir\simeq 2-3\, \zirc$ 
and   goes  to a constant value at large $\zir$, where  $\phi$  becomes constant 
  as it    approaches the minimum of  its 5D potential.
 For a  better understanding of the dilaton effective   potential, 
we can analytically calculate  the effective potential of the 4D tachyon and dilaton in the limit $\zir/\zirc\approx 1$.
  This is   done in  Appendix \ref{appB}.
This shows that the origin of the existence of a  minimum in $V_{\rm eff}(\phi_d)$
can be tracked back to the log-dependence  in \eq{tachy}.

 \subsection{Excitations around the 5D tachyon}

The main  interest of the article is to   know whether close to the conformal transition there is a light dilaton, as often claimed in the literature.
Therefore we will start considering the  flavor-singlet $0^{++}$ spectrum of the theory, to analyze later other sectors.

\subsubsection{The singlet scalar sector and light dilaton}

\begin{figure}[t]
\centering
\hskip-.4cm
\includegraphics[width=1\textwidth]{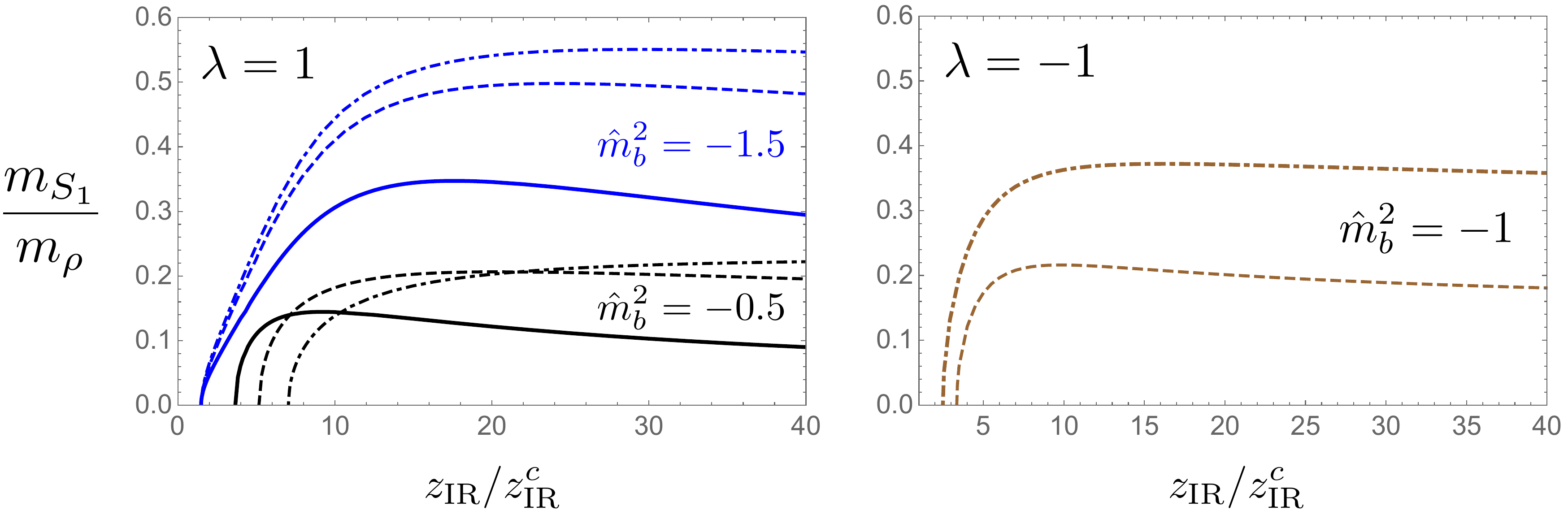}
\caption{\it 
Masses of the lightest  singlet scalar $S_1$ normalized to the vector mass.
We have taken  $\hat\kappa^2=1$ (solid line),  $\hat\kappa^2=4$ (dashed line) and $\hat\kappa^2=10$ (dot-dashed line).
LEFT:  $\lambda=1$ and   $\hat m_b^2=-1.5\ (-0.5)$ for the upper blue  (lower black) lines.
 RIGHT:  $\lambda=-1$ and $\hat m_b^2=-1$.}
\label{plotmassdilaton}
\end{figure}

The  flavor-singlet  $0^{++}$ spectrum is composed by the radion (the only scalar in the gravitational sector) 
and the excitations of  $\Phi_s$   around the background $\phi(z)$. 
Since  the mixing of the dilaton with  $\Phi_s$ is in principle 
 sizable for $N_F\sim N_c$ ($\hat\kappa^2\sim 1$), 
 we must consider  the coupled EOM between the scalar sector and the gravitational sector.
The equations for the  mass spectrum are given in Appendix \ref{appScalar-dilaton} and  must be solved 
 numerically. 

 For  the lightest mode $S_1$ the results are presented in Fig.~\ref{islands} for $\lambda=\pm 1$. 
 We have   normalized  the $S_1$ mass to the one of the lightest vector resonance, $m_\rho$, being this latter the 
lightest state in real QCD and holographic versions \cite{Erlich:2005qh,DaRold:2005mxj}.
Fig.~\ref{islands} shows that  the   $0^{++}$ state is always lighter than the 
 vector in all regions of the parameter space. 
 At the boundary of the regions at which  IR-brane stabilization is achieved,
 the dilaton is massless, but its  mass  is roughly below half of the $\rho$ mass in most of the interior region.
We have checked  that  for larger  $\hat\kappa^2$,  the value of $m_{S_1}$   increases but  not  significantly.

To understand why the dilaton mass  is small,  it is convenient  to show how its mass varies  as a function of $\zir$ for different values of $\hat m^2_b$.  This corresponds to moving in vertical lines in the plane  of Fig.~\ref{islands} from the bottom
 to the top,  trading  the parameter $\delta\hat \Lambda$  for $\zir$ by means of  \eq{minD2}.
We  remark however that we will present  results for a wide region of $\zir$, going beyond  the allowed region \eq{bounds}.
This will help us to understand the origin of the smallness of $m_{S_1}/m_\rho$.
   The result  is  shown in   Fig.~\ref{plotmassdilaton} for different values of $\hat m_b^2>-2$.
    We see  that the dilaton mass starts at zero at $\zir\simeq \zirc$, grows for intermediate $\zir/\zirc$,  
 and tends again to zero for $\zir/\zirc\to \infty$.

We can understand this behaviour analytically.
Assuming that $S_1$ is the radion/dilaton, we can analytically obtain its mass by taking the  derivative of  \eq{minV}  evaluated at the minimum  \eq{minD}. We obtain
\be
m^2_{\phi_d}
\simeq
4\hat\kappa^2 a(\zir)L^2
\left[ \frac{a^2}{24\dot a}  \left(\frac{V_b'\,V_b''}{M_5^2}-V'\right)-\frac{V_b'}{6M_5}
 \right]_{\zir} \partial_{\zir}\phi(\zir)\,,
 \label{md1}
\ee
 where we have used  that at  the minimum $\phi^3_d\partial_{\phi_d}f(\phi_d)/n(\phi_d)\simeq  -4a(\zir)/L$   derived in Appendix~\ref{appA}.
In our particular case \eq{md1} reduces, after   normalizing  to the vector mass \eq{vectormass}, to
\be
\frac{m^2_{\phi_d}}{m^2_{\rho}}\simeq  
P(\phi(\zir))\, Q(\phi(\zir))\, \beta_\phi(\zir)\,,
\label{md2}
\ee
where we have defined the dimensionless functions
\bea
P(\phi(\zir))&\equiv&
\frac{\hat\kappa^2\phi^2(\zir)}{6 m^2_\rho}\simeq
\frac{8\hat\kappa^2L^2\phi^2(\zir)}{27\pi^2\left.\dot A^2_{\rm IR}\right|_{\hat m_b^2=0}}\,,\\
Q(\phi(\zir))
&\equiv&\frac{1}{\dot A_{\rm IR}}\left[\lambda \phi^2(\zir) L^2-(\hat m_b^2+2)^2-4\hat m_b^2 ({\dot A}_{\rm IR}-1)\right]\,,\\
\beta_\phi(\zir)&\equiv&\frac{L}{a(\zir)}\frac{\partial_{\zir}\phi(\zir)}{\phi(\zir)}\,,
\label{betaphi}
\eea
with
\be
\dot A_{\rm IR}\equiv \left. -\frac{\dot aL}{a^2}\right|_{\zir}
=\sqrt{1+\frac{\hat\kappa^2L^2\phi^2}{24}\left(4+{\hat m^4_b}-\frac{\lambda L^2}{2}\phi^2\right)_{\zir}}
\ ,\ \ \  \ \ \dot A_{\rm IR}\geq 1\,,
\ee
where we have used   \eq{wf}.
From \eq{md2} we can infer  different regimes at which  the dilaton can be light:
\begin{itemize}
\item The prefactor $P(\phi(\zir))$  is suppressed for  $\hat\kappa^2 \phi^2(\zir)\ll 1/L^2$.
Therefore in the limit I the dilaton is always light, even when  formally we take     $\hat\kappa^2\gg 1$ (see discussion after \eq{phit}).  Also    for large values of $L\phi$, possible in the limit II with $\lambda<0$,   we have $P\to 1/(L\phi)^2$,
and consequently  the dilaton mass is suppressed.

\item 
The function $Q(\phi(\zir))$ determines the sign of $m^2_{\phi_d}$. 
In the limit I  we have $\phi L\to 0$ and $\dot A_{\rm IR}\to 1$, and  then
$Q$ becomes  negative. 
This   means that  the dilaton effective potential   has actually no  minimum,
as we already pointed out in Section~\ref{sectiond}.
As we increase $\zir/\zirc$,    $Q$  increases till becoming zero,
 corresponding to the points  seen in Fig.~\ref{plotmassdilaton}
with $m_{\phi_d}=0$.
One can check that  
they are inflection points of the  dilaton potential.

\item   The function  $\beta_\phi$   is 
 the main responsible for natural light dilatons in Goldberger-Wise models \cite{GW,Megias:2014iwa}.
Since moving simultaneously the UV and IR boundaries does not change physical quantities,
we can deduce 
\be
\partial_{\zir}\phi(\zir)=-\frac{a(\zir)}{a(\zuv)}\partial_{\zuv}\phi(\zir)\,,
\label{IRUV}
\ee
showing that $\beta_\phi$ is in fact  sensitive to the dependence of the tachyon $\phi$ with  variations of the UV boundary, 
and therefore to the  explicit breaking of conformal invariance (that arises due to  the presence of the UV cutoff).
For this reason $\beta_\phi$ is directly related with  the beta function   $\beta_{\lambda_{\rm eff}}$   of  the dilaton effective coupling of \eq{potdil}.  $\beta_\phi$  explains
why the dilaton mass always goes to zero for large $\zir/\zirc$.
Indeed,  as we approach  the limit II for $\lambda>0$, the 5D tachyon goes to the minimum of its potential
where it becomes constant. We then expect $\beta_\phi\ll 1$.
Also in  the limit II for $\lambda<0$ the slow-roll conditions are achieved
and  $\beta_\phi$  tends to zero.
Unfortunately, these regions of a parametrically  light dilaton are very small in
the full parameter space of the model, 
see Fig.~\ref{islands}, since stabilizing the IR-brane at large $\zir/\zirc$
requires an adjustment of the parameters of the model. 
In the limit I we  can derive from  \eq{eqtachyon} and \eq{phit}  that $\beta_\phi\sim 1/(2\ln(\zir/\zirc))$,
and using the  \eq{minpot} we get   $\beta_\phi\sim \beta$ that is nonzero but smallish in the regions considered.\footnote{Notice that $\partial_{\zir} \phi(\zir)\not= \partial_z\phi |_{\zir}$, and then $\beta_\phi$ does not measure the growth
of the tachyon  (that is power-law $\sim z^2$), but its variation as we move the IR-brane (or UV-boundary) 
that it is much milder (logarithmic).}

\end{itemize}
The situation is similar in regions with $\hat m^2_b<-2$ (see Fig.~\ref{islands} or  Fig.~\ref{spectrum3}).
The only main difference is that the dilaton mass goes to zero  for small values of $\zir/\zircp$, not due to $Q\to 0$,
but because $\phi(\zir)$ tends to a  constant value  as explained in Sec.~\ref{minus2region},
and therefore $\beta_\phi\to 0$.

We  conclude that the dilaton mass  is parametrically smaller than $m_\rho$
at  small and large $\zir$ (respectively corresponding to the left and right boundaries of the regions of Fig.~\ref{islands}). 
At small $\zir$, 
the reason is either the existence of an inflection point in the dilaton potential  (for the case $\hat m^2_b>-2$)
or  that $\phi(\zir)$ becomes  frozen  and $\beta_\phi\to 0$ (for the case $\hat m^2_b<-2$).
Also at small 5D tachyon values  its  log-dependence  on $\zir$ gives a smallish $\beta_\phi$
and therefore a smallish dilaton mass.
At large $\zir$  (limit II) the geometry approaches again   AdS$_5$  (the dual model flows towards another approximate CFT$_4$)
where scale invariance is partially recovered and therefore the dilaton mass must go to zero.
"Trapped" between these two limits,  the dilaton mass  
cannot  grow much  in the intermediate region  
and then  remains always the lightest resonance (although not parametrically lighter than the others).

Finally, we would also like  to discuss the mass of the  second lightest singlet scalar, $S_2$.
This is also obtained numerically  (see Appendix~\ref{appScalar-dilaton}), 
and the result is shown  in Fig.~\ref{spectrum1} for certain representative values of the parameter space.
This scalar $S_2$ is mostly   the excitation around the profile $\phi(z)$ (up to a small mixing with the radion), a Higgs-like state.
For this reason  when  $\zir\to \zirc$, 
we expect   $m^2_{S_2}\to \lambda_t \phi_t^2\to  0$,
as appreciated in Fig.~\ref{spectrum1}.

\subsubsection{Non-singlet scalars, vector and axial-vector excitations}

For the   scalars in the  adjoint  under the $SU(N_F)_V$ symmetry,
$\Phi_a$,
the EOM  is given by 
\be
[\partial_\mu^2-a^{-3}\partial_z a^3\partial_z
+a^2 M_{\Phi}^2+a^2(3\lambda-2N_F\lambda_2) \phi^2(z)]\Phi_a=0\,.
\label{seomadj}
\ee
As we already mentioned, there are two important difference with respect the singlet scalar case. First,
the scalars in the adjoint do not mix with the radion/dilaton.  Second, the quartic
coupling in \eq{seomadj} is different from the singlet case  due to the presence of  $\lambda_2$.
This implies that the adjoint scalar masses are expected to be different from the singlet scalar masses,
with the magnitude of the  mass splitting being sensitive to $\hat\kappa^2$ and $\lambda_2$.

We are  also interested in  the vector $V_{M}=(L_M+R_M)/\sqrt{2}$ and axial-vector $A_{M}=(L_M-R_M)/\sqrt{2}$ spectrum \cite{Erlich:2005qh,DaRold:2005mxj}. 
The vector spectrum is only indirectly sensitive to the tachyon  through its impact to  the metric.
Therefore, since    flavor-singlet  resonances (the $\omega$ in QCD) 
and  adjoint   resonances (the $\rho$ in QCD)   feel the same metric and have the same boundary conditions,
 they get equal masses.
 This is an important prediction of the 5D model.\footnote{
Of course,  mass splittings could be generated at the loop level or from
higher-dimensional operators in \eq{la5}, but these  are expected to be suppressed.
 }

It  is useful to have an approximate analytic value for $m_\rho$, since we are using  this mass  to normalize the other resonance masses. This is possible  in the limit in which the 5D space  is approximately  AdS, that corresponds 
to  limits I and II, as we explained in Section~\ref{tasol}.  In AdS$_5$ we have
$m_\rho\sim -({3\pi}/{4})({\dot a}/{a})$. 
 We find that a reasonably good approximation for ${\dot a}/{a}$ 
   in the limit of small and large $\zir$   is given by  \eq{wf} 
   neglecting the derivative terms    and  taking $\phi$ at $z=\zir$.\footnote{We put to zero the derivative terms to avoid the drastic change  of $\phi$ near the IR boundary -see footnote~\ref{footy}.} 
   We then have: 
\be
m_\rho\simeq \frac{3\pi}{4}\frac{a(\zir)}{L} \left.\dot A_{\rm IR}\right|_{\hat m_b^2=0}\,.
\label{vectormass}
\ee
We have checked that this value  is within $\lesssim 20\%$ the exact mass of $\rho$ for the regions of the parameter space studied in this article.

 \begin{figure}[t]
\centering
\hskip-.4cm
\includegraphics[width=0.68\textwidth]{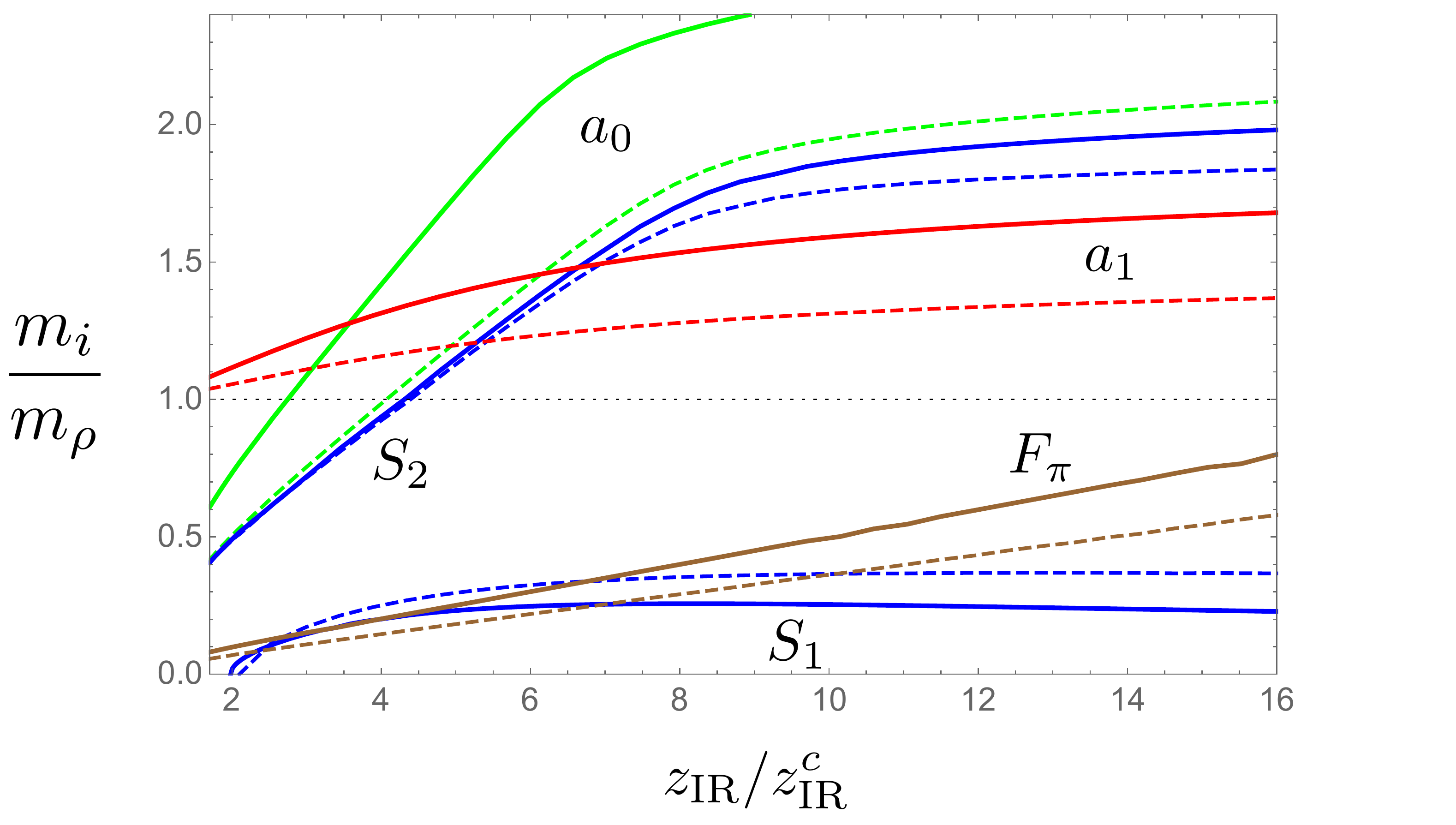} \ \
\caption{\it 
Masses of the two lightest  singlet scalars, $S_1$ and $S_2$, lightest   adjoint scalar ($a_0$),
lightest   axial-vector ($a_1$) and $F_\pi$, normalized to the vector mass for  $\hat m_b^2=-1$,  
$\lambda=1$, $\lambda_2=-2$, $g^2_5=1$ and  $\hat\kappa^2=1\ (4)$ for the solid (dashed) line.
}
\label{spectrum1}
\end{figure}

The axial-vector spectrum   depends directly  on the $\phi$  profile via \eq{la5s}, being this responsible for the mass splitting from the vector spectrum. 
Another important quantity is   the  Goldstone decay constant $F_\pi$, that is the order parameter of the
chiral breaking.
This can be calculated via holography from the
  axial-vector two-point correlator at zero momentum \cite{Erlich:2005qh,DaRold:2005mxj}:
\be
F_\pi=\Pi_A(0)=-\frac{M_5L}{2\zuv}\left.\frac{ \partial_z  A(z)}{A(z)}\right|_{\zuv}\,,
\ee
where $A(z)$ is the 5D solution of the axial-vector with Dirichlet UV-boundary condition.

The results (with no approximations) are   shown in Fig.~\ref{spectrum1} for  some representative values of the parameter space.
Following the notation in QCD, we denote with   $a_0$ and $a_1$  the adjoint scalar and axial-vector respectively.
Since $F_\pi$ is the only quantity that depends on $M_5$ ($N_c$ in the dual theory),
we have fixed its value using \eq{fit} with $N_c=3$.
For  $\zir\approx\zirc$ (limit I) where the chiral breaking is small,
we see that indeed   $F_\pi$ and  $(m_\rho-m_{a_1})/m_\rho$ are small.
As we increase $\zir$, we move towards limit II where the breaking 
of the chiral symmetry  is larger, as can be appreciated by the growth of $F_\pi$ 
and    $S_2-a_0$ and   $\rho-a_1$ mass splittings.
On the other hand, the mass of $a_0$ strongly depends on $\lambda_2$, 
and we have chosen  a negative value, $\lambda_2=-2$,  that makes  the mass splitting with the singlet sector
positive, as lattice simulations (see later) seem to suggest.
 Similarly to the singlet scalars, we also have  that $m_{a_0}$ goes to zero as $\zir\to\zirc$,  since 
 the tachyon value  goes to zero in this limit and we recover the chiral symmetry.

Let us briefly comment on what  happens for other values of the parameters of the model.
The effect of $\hat\kappa^2$ in the mass spectrum  is clear from Fig.~\ref{spectrum1}
where we show the spectrum for two different values of $\hat\kappa^2$. 
The main effect is that  as we increase $\hat\kappa^2$, the profile of $\phi$ becomes flatter and  smaller,
as appreciated in Fig.~\ref{tachyonprofile}, giving  a smaller breaking of the chiral symmetry.
The spectrum is  mildly  sensitive to the values of  $\hat m^2_b$, unless we take 
$\hat m^2_b< -2$ that we will discuss later (Fig.~\ref{spectrum3}).
Finally,  $g_5^2$ only affects $F_\pi$ and $m_{a_1}$  that will increase as  $g_5^2$ increases.

\begin{figure}[t]
\centering
\hskip-.4cm
\includegraphics[width=1.02\textwidth]{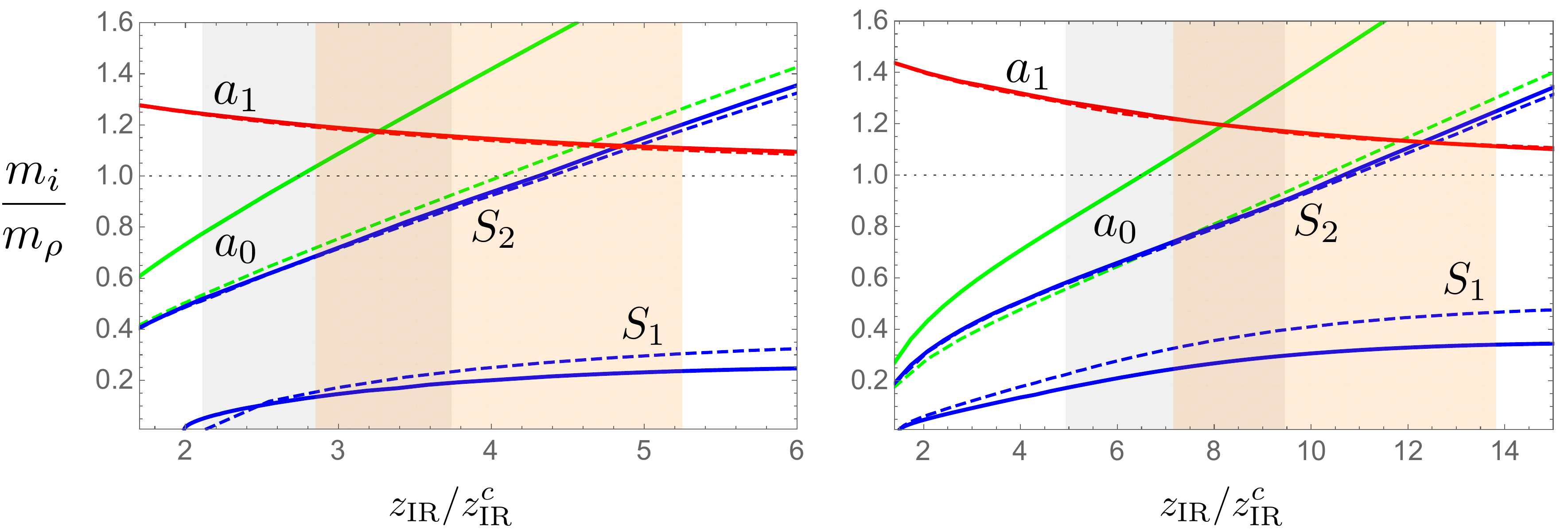}
\caption{\it 
Masses of the two lightest  singlet scalars, $S_1$ and $S_2$, lightest   adjoint scalar ($a_0$),
lightest   axial-vector ($a_1$) normalized to the lightest vector mass ($m_\rho$)
 for constant $F_\pi\sim m_\rho/7$
 as a function of $\zir/\zirc$ for    $\lambda=1$.
  We have taken   $\hat\kappa^2=1\ (4)$ for solid (dashed) lines.
  The left grey (right orange) band corresponds to the region $0.5<g^2_5<2$  for $\hat\kappa^2=1\ (4)$.
  LEFT:  $\hat m_b^2=-1$. RIGHT:  $\hat m_b^2=-1.5$.}
\label{spectrum2}
\end{figure}

It is  more instructive, also in part  to compare later our results with lattice simulations,
to analyze the spectrum  at constant $F_\pi$.
 For this purpose, we adjust  $g^2_5$ to  fulfill,  for the different values of $\zir/\zirc$ (or equivalently $\delta\hat \Lambda$),   the relation $F_\pi\simeq m_\rho/7$ as in QCD.  
The results are given in   Fig.~\ref{spectrum2} for $\hat m^2_b=-1$ (left) and $\hat m^2_b=-1.5$ (right).
We have kept $g^2_5$ in the interval $0.5<g^2_5<2$ and this has limited
the possible values of $\zir/\zirc$  to the  blue and orange bands for $\hat\kappa^2=1$ and $4$ respectively.\footnote{The constraint  $\delta \hat\Lambda\leq 0$ has not been imposed. If we impose it, 
we obtain  $\zir/\zirc\geq  3.06\ (3.37)$  for $\hat\kappa^2=1\ (4)$ in the left plot of Fig.~\ref{spectrum2},
and $\zir/\zirc\geq 1.88\ (1.9)$ for the right plot.}
The main conclusions from Fig.~\ref{spectrum2} are the following. 
The lightest resonance is always the scalar $S_1$,  a dilaton-like state.
The $S_2$, the Higgs-like state, is also smaller or around  $m_\rho$, 
and can only be larger if we take large values of $\zir/\zirc$
(that implies small values of $g_5^2$ in order to keep $F_\pi\sim m_\rho/7$).
The ratio  $m_{a_1}/m_\rho$ is closer to $1$ than in real QCD where $m_{a_1}/m_\rho\sim 1.6$
or previous holographic QCD versions  \cite{Erlich:2005qh,DaRold:2005mxj}.
The mass of $a_0$ is also smaller than in real QCD.
 As we will see in the following, these  properties are also found in lattice QCD for large $N_F$. 

 \begin{figure}[t]
\centering
\hskip-.4cm
\includegraphics[width=0.6\textwidth]{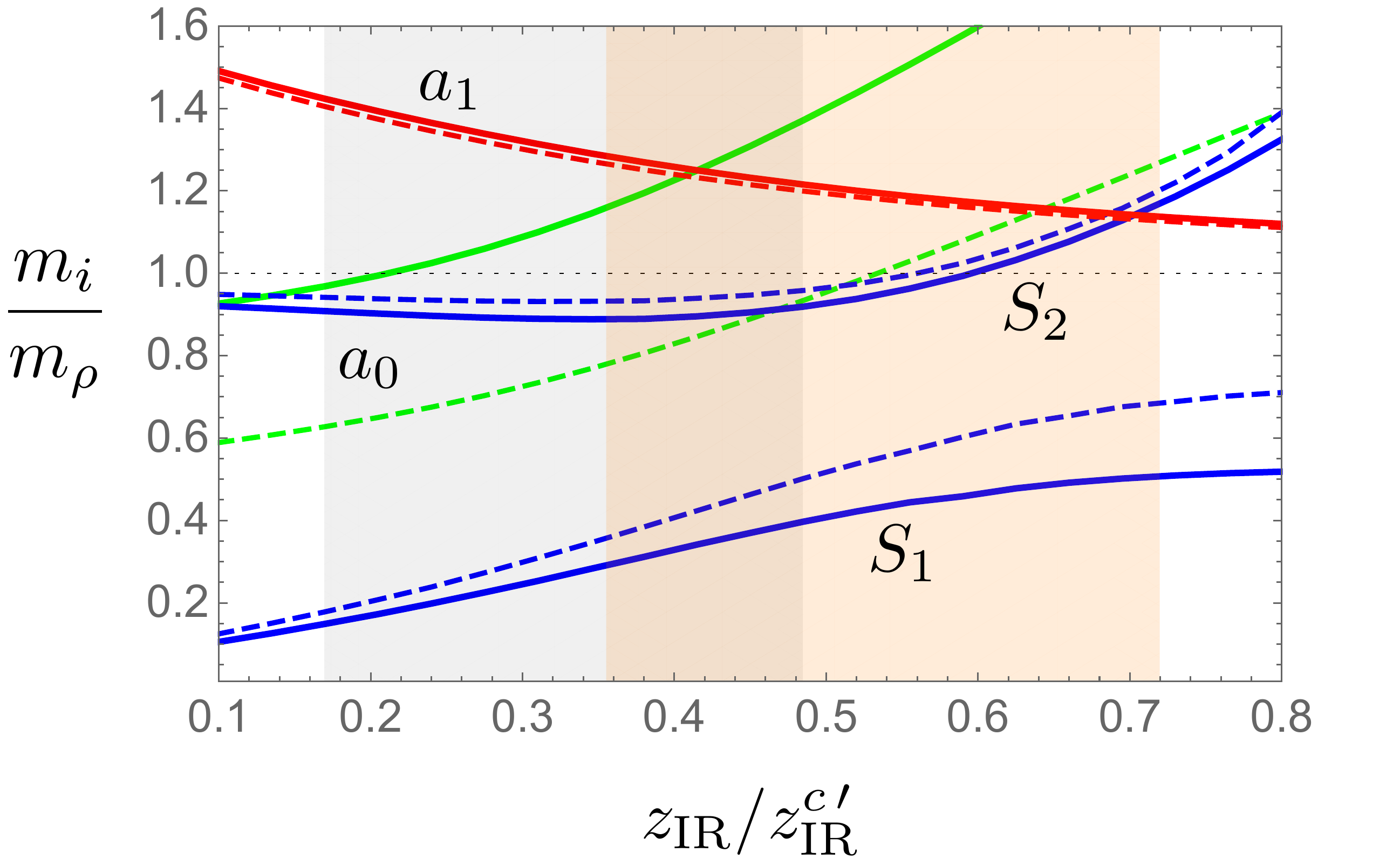} \ \
\caption{\it 
As in Fig.~\ref{spectrum2} but for   $\hat m_b^2=-3$.}
\label{spectrum3}
\end{figure}

The situation is only slightly modified in the region $\hat m^2_b<-2$.
In Fig.~\ref{spectrum3} we show the mass spectrum for $\hat m^2_b=-3$.
The main differences with respect  Fig.~\ref{spectrum2} is in the scalar mass spectrum where we appreciate that  at smaller values of $\zir/\zircp$, where here $\zircp\equiv e^{\pi/\sqrt{\epsilon}}\zuv$,  the $a_0$ and $S_2$ masses do not go to zero.
The reason is the following. As explained in Section~\ref{minus2region}, for $\hat m^2_b<-2$ 
 the profile of  $\phi$ is always non-zero (unless   $\zir<\zirc\sim \zuv$).
This implies that we do not recover the chiral symmetry in the region of interest, $\zir\sim \zircp$, and  the $S_2$ and $a_0$
 masses never approach  zero. Nevertheless, their masses are predicted to be  around the $\rho$ mass.

The main lesson that we have learned    on the mass spectrum of $S_2$, $a_0$ and $a_1$    
is that they seem to tend to be lighter in models close to the  conformal  transition 
(as compared to real QCD that is far from the conformal  critical point).
 What is  the reason for that?
  As it is well-known, the  dimension of a scalar operator has a minimal value  determined by its unitarity bound,  in this case 
 Dim$[{\cal O}_*]=1$, a limit at which the scalar decouples from the CFT \cite{Grinstein:2008qk}. 
 Therefore it is expected that, as a scalar operator approaches  this decoupling limit,  the mass of the lightest resonance
 associated to it  becomes smaller.
By using the AdS/CFT correspondence this means, via \eq{dic1},  that the lightest $\Phi$ resonance is expected to be lighter
the more we approach the BF-bound.
We  can also understand this  ``geometrically".
The wave-function of the  lightest scalar grows as $z^{2+\sqrt{4+M^2_\Phi L^2}}$, 
that implies that the wave-function becomes flatter
and spread more into the AdS$_5$ space as we approach the BF-bound $M^2_\Phi L^2\to -4$.  
In this limit, then,   the  scalar excitation becomes less sensitive to the IR and therefore its mass is expected to be smaller (see also~\cite{Vecchi:2010em}).\footnote{We could make the wave-function even flatter by quantizing differently the scalar following  Ref.~\cite{Klebanov:1999tb}
(valid for $-4<(M_\Phi L)^2<-3$). 
In this case we have ${\rm Dim}[{\cal O}_*]=2-\sqrt{4+M^2_\Phi L^2}$ that means that the scalar is the dual of an operator of 
dimension between $1$ and  $2$. 
We will get in this case the wave-function  $z^{2-\sqrt{4+M^2_\Phi L^2}}$, reaching the full decoupling from the IR (from the CFT) at $M^2_\Phi=-3/L^2$ when  the mode becomes non-normalizable. Nevertheless
the scalar   becomes UV  sensitive and it is not expected to survive in the spectrum.}

The fact that the profile of  $\phi$ becomes flatter as we approach the BF-bound  also explains
the smaller mass splitting between $a_1$ and $\rho$ than in real QCD. 
Indeed, if we keep $F_\pi$ constant,
the flatter the $\phi$ profile, the smaller  $\phi(\zir)$.
Since the $a_1$ wave-function is peaked towards the IR-brane, it is  mostly sensitive to the value of 
$\phi(\zir)$. Therefore, as the 5D mass of $\phi$ gets closer to the BF-bound, we  expect $m_{a_1}$ to be less sensitive to
chiral breaking. For the same reason we understand  $m_{a_1}$ becoming  smaller as we increase $\zir$ (see Fig.~\ref{spectrum2}), as    $\phi$ becomes   flatter for larger $\zir$.

\section{Comparison with Lattice QCD in the large $N_F$}
\label{comparison}

Lattice results for QCD with $N_F=8$ have been reported in Ref.~\cite{Aoki:2016wnc,Brower:2015owo}.
At such  large value of $N_F$, it is believed that QCD is close to the conformal transition,  expected to occur
around $N_F\sim 9$.
It was found \cite{Aoki:2016wnc,Brower:2015owo}
\be
\frac{F_{\pi}}{m_\rho}\simeq 0.14\ ,\ \ \frac{m_{f_0}}{m_\rho}\simeq 0.5\ ,\ \  \frac{m_{a_0}}{m_\rho}\simeq 1\ ,\ \  \frac{m_{a_1}}{m_\rho}\simeq 1.4\,,
\label{lattice}
\ee
where $f_0$ is the lightest flavor-singlet $0^{++}$ state ($S_1$ in our notation).
It is instructive to compare them with real QCD that is supposed to be far  from the conformal edge.
We have \cite{Patrignani:2016xqp}
\be
\frac{F_{\pi}}{m_\rho}\simeq 0.13\ ,\ \ \frac{m_{f_0}}{m_\rho}\simeq 1.3\ ,\ \  \frac{m_{a_0}}{m_\rho}\simeq 1.3\ ,\ \  \frac{m_{a_1}}{m_\rho}\simeq 1.6\,.
\label{realqcd}
\ee
We see that close to the conformal transition, we spectrum of  \eq{lattice} shows,  as compared to real QCD \eq{realqcd},  lighter $f_0$ and $a_0$ scalars,  and a smaller mass splitting  between the $\rho$ and $a_1$ resonance. 
Surprisingly, the ratio of $F_\pi/m_\rho$ is quite similar to real QCD, showing that this quantity
is quite independent of $N_F$.

Let us compare our results to the values of \eq{lattice}. 
In order to reduce the number of parameters, we can match  the predictions of our model at the UV with  those of QCD with $N_F$ flavors. In particular,  the two-point vector-vector  correlator at large momentum $p^2$ is given in our model by
\cite{DaRold:2005mxj} 
\be
\Pi_{V}(p^2)\simeq -\frac{M_5L}{2g_5^2}p^2\ln(p^2\zuv^2)\,,
\ee
that matching to that of QCD with $N_F$ flavors gives
\be
\frac{M_5L}{g^2_5}=\frac{N_c}{12\pi^2}\,.
\label{fit}
\ee
Using \eq{fit} our predictions for the mass spectrum were presented  in   Fig.~\ref{spectrum2} and \ref{spectrum3}
for  $F_\pi=m_\rho/7$.
We see that  our predictions  on the spectrum of resonances follow quite close  the pattern  \eq{lattice}.
We have $m_{a_1}/m_\rho$ closer to one than in QCD,
with the scalars $a_0$  and $S_1$ being lighter than the $\rho$ in most of the parameter space.
Indeed,  in the region   $1.2<m_{a_1}/m_\rho<1.4$, we  find $m_{a_0}/m_\rho\lesssim 1$
and  $m_{S_1}/m_\rho\lesssim 0.3$.

There are other important     predictions arising from  our  holographic model 
that would be interesting to check   in future lattice simulations.
For example, as we already mentioned, the  mass splittings between the adjoint and singlet  vectors
 is zero at  leading order, and can only arise  from loop effects or higher-dimensional operators that are suppressed. Also the second singlet scalar $S_2$ (a Higgs-like scalar) seems to be  lighter than the $\rho$ in
the region where $1.2<m_{a_1}/m_\rho<1.4$. 
Finding this second resonance so light  would be a clear indication that the lightest scalar $S_1$ is a dilaton and not a Higgs-like state. Other properties of the scalars, such as decay constants or couplings, that can also be calculated in these holographic  models, are left  for future work.

\section{Models for  the hierarchy problem}
\label{hierarchy}

The model described here open new possibilities for generating small scales. Since  the IR-brane is naturally stabilized
at ${\zir}\sim O( {\zirc})$, we have a way to generate  exponentially small scales.
Indeed, from \eq{zirc} we have
\be
\frac{1}{\zir}\sim \frac{1}{\zirc} = e^{-\pi/\sqrt{\epsilon}}\frac{1}{\zuv}\ll \frac{1}{\zuv}\,.
\label{hiera}
\ee
The presence of a scalar with a mass just below the BF bound can also be achieved dynamically.
If the mass of $\Phi$ is  $z$-dependent, for example,
$M^2_\Phi L^2=-4-{\cal E}(z)$, where ${\cal E}(z)$ slowly varies 
 from negative to positive values as $z$ increases,
the mass of $\Phi$ will cross  the BF bound at the position $z=\zuv'$ at which ${\cal E}(\zuv')=0$.
For example, we can  consider ${\cal E}(z)=\epsilon \ln (z/\zuv')$ with $\epsilon\ll 1$
(for other cases, see \cite{Hubisz}).
This $z$-dependent mass for $\Phi$  
can be easily achieved by promoting ${\cal E}(z)$ to a scalar $R$ with a 5D potential  
$V= -\sqrt{\epsilon} R(1+|\Phi|^2/L^2)$. 
 This  scalar gets a profile $R(z)=\sqrt{\epsilon} \ln (z/\zuv)$, giving a contribution to the mass of  $\Phi$
 proportional to $\epsilon \ln (z/\zuv)$.

For   $M^2_\Phi L^2=-4-{\cal E}(z)$ with  ${\cal E}(z)=\epsilon \ln (z/\zuv')$,
the wave-function of the massless mode is   not anymore  \eq{eqtachyon} but
\be
\phi(z)=\frac{\phi_t(x)}{N}\, z^2 \sqrt{\ln({z}/{\zuv})}\,  J_{1/3}\left(\frac{2}{3}\sqrt{\epsilon} \ln^{3/2}\frac{z}{\zuv}\right)\,,
\label{eqtachyon2}
\ee
where $J_{1/3}$ is a Bessel-function of order $1/3$, and the IR-boundary condition  \eq{bc} at $\zir=\zirc$ leads  now in the limit 
$\epsilon\to 0$ to
\be
\frac{2}{3}\sqrt{\epsilon} \ln^{3/2}\frac{\zirc}{\zuv}\simeq\left(n-\frac{1}{12}\right)\pi\ ,\ \  n=1,2,...\,,
\label{zirc2}
\ee
corresponding to the zeros of the Bessel function.
The situation is quite  similar to the case of constant ${\cal E}(z)$ discussed above;
the only important difference worth to mention is that in the limiting case II with $\lambda>0$,
the maximal value of $\phi$ is not constant, 
as $M^2_\Phi$ evolves logarithmically.  The theory has evolved into  a deformed CFT.  

We leave the implications of these scenarios for the electroweak scale for future work.
We only point out several interesting features. 
First, 
the lightness of the dilaton can have important implications for the LHC \cite{LHC}.
Also the fact that the mass of $a_1$ is closer to the mass of $\rho$ 
implies smaller values for  the $S$-parameter, as favored by precision experimental data.
Furthermore,  having the operator that drives symmetry breaking 
a dimension close to $2$
helps to pass  flavor constraints \cite {Panico:2016ull}.
Also   it was  shown 
in Ref.~\cite{Baratella:2018pxi} that these models can lead to a long period of supercooling
in the early universe with implications in Dark Matter and axion cosmological abundances. 

\section{Conclusions}
\label{conclusions}

We have used holography to study  strongly-coupled theories close to the conformal transition, that is
the transition from  the  non-conformal regime to   the conformal one.
This  transition is expected to happen in gauge theories (such as QCD) as the number of fermions $N_F$ increases.
Recent lattice results \cite{Aoki:2016wnc,Brower:2015owo}
have shown that as we get closer to the conformal transition, the lightest resonance
is a  $0^{++}$ state, claimed to be a dilaton.

We have followed the idea of Ref.~\cite{Kaplan:2009kr} that suggested that conformality is lost when  the IR fixed point merges with a UV fixed point,   as shown in Fig.~\ref{merging}.
Holography tells  that this must occur by  an operator ${\cal O}_*$ (probably $q\bar q$ in QCD) 
whose dimension is equal to two    that gets a small imaginary part  when leaving the conformal regime.
In the gravitational dual models this is driven by a  scalar whose  mass goes below the BF bound and becomes tachyonic.

We have presented a very simple extra-dimensional model with the essential ingredients to study the conformal transition and calculate the mass spectrum. The model  consists  of a five-dimensional gravitational sector with a scalar and gauge bosons associated to the global $SU(N_F)_L\otimes SU(N_F)_R\otimes U(1)_B$. We have  allowed  for the most general Lagrangian 
following  the 5D EFT rules, and explained the connection between the 5D couplings and 
 the large $N_c$ and $N_F$ expansion.
  To  model confinement we  cut off the space by  an IR-brane that we showed to be stabilized by the presence of the tachyon.

We have calculated the mass spectrum of this 5D model, 
showing that  indeed the  dilaton   corresponds  to the lightest resonance.  
To understand this property, we have derived a simple formula  for the dilaton mass,  \eq{md2}.
This shows that the mass  of the dilaton crucially  depends on $\beta_\phi(\zir)$  given in \eq{betaphi}
that is sensitive  to the  variation of  $\phi(\zir)$  as we move the UV boundary 
(therefore sensitive to the explicit breaking of the conformal symmetry).
Either for small or large values of $\zir$, we have shown that $\beta_\phi(\zir)\to 0$ 
and therefore the dilaton mass tends to zero.
For small $\zir$ this is due to either the existence of an inflection point in the dilaton potential  (for $\hat m^2_b>-2$),
or  that $\phi(\zir)$ becomes  constant  (for $\hat m^2_b<-2$).
Also for small  $\phi(z)$, where we can perform analytical calculations, 
we find a mild log-dependence  of $\phi(\zir)$ with $\zir$ (and therefore a smallish $\beta_\phi(\zir)$) that can be traced back to  
the explicit breaking of the conformal symmetry due to the double-trace marginal operator 
${\cal O}_g=|{\cal O}_*|^2$.
 For large $\zir$,  also $\beta_\phi(\zir)\to 0$  as the tachyon either goes to the minimum of its potential and becomes constant or enter into a "slow-roll" condition, meaning that the geometry approaches again   AdS$_5$.
In between these two limiting cases, the dilaton can become heavier but its mass cannot grow enough to overcome  $m_\rho$.
Therefore the dilaton is found to be lighter than the rest of the resonances, although  it is never parametrically lighter
 in most of the area of the allowed parameter space, as shown in Fig.~\ref{islands}.

We have compared our predictions with lattice results for QCD with a large $N_F$ (\eq{lattice}) 
and showed that  our model predicts quite similar resonance mass pattern: the lightest state is the singlet $0^{++}$,
with the adjoint scalar $a_0$  mass close to $m_\rho$ and lighter than in real QCD.
We have also shown than the   mass splitting between the vector ($\rho$) and axial-vector ($a_1$) is smaller 
close to  the conformal transition.  We have given a geometric explanation for these properties.
Furthermore, the 5D model proposed here also  provides extra  predictions that lattice could check in the future.
For example,  we find that the  second $0^{++}$ state,  $S_2$,   is mostly a Higgs-like state ($q\bar q$ state) 
with  a mass  around $m_\rho$, similarly as $a_0$.
The 5D model also predicts that the masses of the flavor singlet and  adjoint  vector resonances are  similar 
(as it happens also in real QCD).

There are several interesting calculations that are left for the future.
For example, it is also possible to calculate decay constants and couplings of the resonances 
along the lines of Ref.~\cite{DaRold:2005mxj}.
One could also   easily  add  explicit quark masses to the model to see the impact on the spectrum,
or study the model at the conformal edge but inside the conformal window.
It could also be interesting  to understand what are the  holographic versions of  
the  complex CFT described in Ref.~\cite{Gorbenko:2018ncu}.
Finally, as discussed above, this type of models can provide a new approach to the   hierarchy problem 
with a clear  impact on  LHC phenomenology as the  $0^{++}$ resonance is expected to be the lightest one.
All these issues clearly deserve more attention.

\section*{Acknowledgements}
\label{sec:acknowledge}

We would like to thank Eugenio Megias, Giuliano Panico and Mariano Quiros.
AP  was supported   by the Catalan ICREA Academia Program.
LS was supported by a Beca Predoctoral Severo Ochoa del Ministerio de
Econom\'{\i}a y Competitividad (SVP-2014-068850).
This work was also partly supported by the grants FPA2017-88915-P, 2017-SGR-1069 and 
SEV-2016-0588.

\appendix

\section{Scalar and gravity coupled equations of motion}
\label{appA}

In this appendix we present  the  equations of motion (EOM) of the scalar and gravitational sector,
that we use in this article in  order to derive the background and  mass spectrum of the model.
For this purpose it is useful   to  work with proper coordinates, as the   metric-scalar system of EOM
simplifies.
Once the results are obtained, 
we have rewritten them  in conformal coordinates \eq{metric} to be presented in the main text.
Conformal coordinates  allow   a better interpretation of the results as $1/z$ determines  the natural mass   scale at the position $z$.

\subsection{Scalar-metric system} 

In proper coordinates $\{x^\mu,y\}$
 the background metric can be written as
\begin{align} \label{A.B1}
ds^2=e^{-2A(y)}\eta_{\mu\nu}dx^{\mu}dx^{\nu}-dy^2\,,
\end{align}
where $\eta_{\mu\nu}=\diag(1, -1, -1, -1)$,  $0\leq y\leq y_{\rm IR}$  with 
the IR-brane  localized at $y=y_{\rm IR}$, and we have  conveniently  rewritten the warp factor as $a=e^{-A}$. 
The 5D EOM for the metric-scalar system,  that follow from  the action in \eq{s5} in these coordinates,  are given by 
\begin{align} 
\ddot{\phi}&=4\dot{A}\dot{\phi}+V'\,,\quad \label{A.B2} \\
\dot{A}&=\sqrt{\f{1}{L^2}+\f{\hat\kappa^2L^2}{12}\Big(\f{\dot{\phi}^2}{2}-V(\phi)\Big)}\,,\label{A.B3} \\
\ddot{A}&=\f{\hat\kappa^2L^2}{6}\dot{\phi}^2\,,\label{A.B4}
\end{align}
where in this appendix   $\dot{\phi}\equiv\partial_{y}\phi$, $\dot{A}\equiv\partial_{y}A$. At the IR-brane
we must impose  the  IR-boundary conditions:
\begin{align}\label{A.B5} 
&\left(M_5\dot{\phi}+V'_{b}(\phi)\right)\Big|_{y_{\rm IR}}=0\,, \\
&\left(\f{6M_5}{\hat\kappa^2L^2}\dot{A}+V_{b}(\phi)\right)\Big|_{y_{\rm IR}}=0\,,
\end{align}
where the second equation is the junction condition that determines the value  $y_{\rm IR}$ where the IR-brane is dynamically stabilized. 
Plugging \eq{A.B3} into \eq{A.B2} gives a differential equation  involving  only $\phi$  that can be easily solved.
Afterwards,  we can solve \eq{A.B3} to obtain the metric warp factor $A(y)$.
We can  go  to conformal coordinates
by using ${dy}/{dz}=e^{-A(y)}$.

Working with proper coordinates, the slow-roll conditions are, in analogy with inflation,   given by
\begin{align}\label{SR1} 
\f{\dot{H}}{H^2}\ll 1
\quad \hbox{and} \quad
\f{1}{H}\f{\partial_y\Big[\f{\dot{H}}{H^2}\Big]}{\Big[\f{\dot{H}}{H^2}\Big]}\ll 1 \,,
\end{align}
where $H=\dot{A}$. Using  \eq{A.B2}--\eq{A.B4}, the slow-roll conditions \eq{SR1} can be written in the following  equivalent form:
\begin{align}\label{SR2} 
\f{(V')^2}{(V-\f{12}{\hat\kappa^2L^4})^2}\ll\hat\kappa^2L^2
\quad \hbox{and} \quad
\f{V''}{V-\f{12}{\hat\kappa^2 L^4}}\ll \hat\kappa^2L^2\,.
\end{align}
Since  we work with polynomial potentials,  
the two slow-roll conditions \eq{SR2} reduce to one condition
when the feedback of the metric becomes important $(V\gg\f{12}{\hat\kappa^2L^4})$. This is given by
\begin{align}\label{SR3} 
\hat\kappa^2 \phi^2 \gg 1/L^2\,.
\end{align}

\subsection{Singlet scalar-dilaton system}
\label{appScalar-dilaton}
If the dilaton is not a priori assumed to be light, 
we must solve exactly the eigenmasses of the scalar sector considering the mixing  between the singlet scalar $\Phi_s$ and the dilaton, which is of order one for $N_F\sim N_c$. 
This is done conveniently in  a diagonal gauge where the brane is straight,
corresponding to a constant value of the extra coordinate $y_{\rm IR} = $const. In this gauge, the EOM
 reduces to  Eq.~(3.17) of Ref.~\cite{Csaki:2000zn}, that we can write as
\begin{align}\label{A.B6}  
{\cal D}\psi_n&=\f{3m_n^2}{\hat\kappa^2 \dot{\phi}^2L^2} e^{2A}\psi_n \quad \hbox{with} \quad
{\cal D}\equiv 1-\partial_y \Big[ \f{3e^{2A}}{\hat\kappa^2 \dot{\phi}^2L^2} \partial_y[e^{-2A}] \Big]\,,
\end{align}
with the IR-boundary condition:
\begin{align} 
m_n^2 \psi_n\Big|_{y_{\rm IR}}=\Big(V''_{b}+\f{\ddot{\phi}}{\dot{\phi}}\Big) \partial_y[e^{-2A} \psi_n]\Big|_{y_{\rm IR}}\,.
\label{A.B7}  
\end{align}

\subsubsection{Light dilaton limit}

When the dilaton becomes  the lightest mode of the scalar sector, we can analytically derive its mass, as given  in \eq{md1}. Here we present the details to obtain this mass. 

The physical meaning of the dilaton field is the IR scale that appears dynamically in the theory. In the 5D model, this is incarnated geometrically by the location of the IR-brane which is indeed dynamical. The picture is more transparent by allowing the IR position to be $x^\mu$-dependent, {\it i.e.}, that it is a 4D field. This is equivalent to using a gauge where the IR-brane location is not straight, but rather  defined by the surface $y=y_{\rm IR}(x^\mu)$. On the other hand, the variable that transforms under a scale dilatation of the $x^\mu$ coordinates is not the proper coordinate $y$ itself but the warp factor $a(y)$. Therefore it is natural to identify the dilaton field with the warp factor evaluated at the IR-brane location,
\be\label{phiDy}
\pd = \frac{1}{L} e^{-A(y_{\rm IR})} \,.
\ee
This variable is also convenient because it is easy to extract the normalization of both the potential and of the kinetic terms in terms of it.  For instance, with this definition 
the brane tension term (proportional to $\sqrt{-g^{IR}}$ with $g^{IR}_{\mu\nu}$ the induced metric on the brane) is simply a quartic coupling, $\pd^4$. 
Below we will see that this variable is actually not canonically normalized in general, although its kinetic term can be easily found. From the normalization of both kinetic and potential terms then a formula for the mass will follow. 

Let us start by the potential term. We can redo the argument around \eq{minV} in terms of $\pd$, that allows to reconstruct quite directly the derivative of the off-shell effective potential, $dV_{\rm eff}/d \pd$, which must be proportional to ${\pd}^{\,3}$  and to the junction condition.  The overall normalization constant  can be fixed by requiring that in the $\hat\kappa^2\to0$ limit
the effective potential reduces to 
 $V_{\rm eff} = (M_5\,L)\int_{bulk}  (V+\dot\phi^2/2) + V_b$, with  
$\phi$ solving the EOM and  depending parametrically  on $y_{\rm IR}$ .
This leads to
\be
\frac{dV_{\rm eff}(\pd)}{d\pd}=24 
\left.
\frac{M_5}{\kappa^2} \,
\big( L \, \pd \big)^3
\,\left(
\sqrt{1+\frac{\hat\kappa^2L^4}{12}\Big(\frac{V_b^{'\, 2}}{2M^2_5}-V\Big)} \,
+ \,\frac{\hat\kappa^2 L^3}{6M_5}\, V_b\right)\right|_{y_{\rm IR}}\,,
\label{minV2}
\ee
that differs from \eq{minV} by an overall multiplicative constant;
  this does not matter much however for the 
mass \eq{md1} as long as we factor out the same constant in the kinetic term.

Next, the normalization of the dilaton kinetic term. Another advantage of using a non-straight gauge is that all the kinetic term contributions arise only from localized terms on the IR-brane itself. This is welcome because the dilaton is an IR mode and its properties should arise from the IR only. Moreover, it is also convenient because it allows to identify these kinetic term contributions in the `probe' limit, where we ignore how the brane bending sources the 5D metric. 
%
%
Following \cite{Garriga:2001ar}, one quickly sees that the radion/dilaton kinetic term arises from two sources.  First, the brane tension (potential) term, via the determinant of the induced metric on the brane, 
\be
\sqrt{-g^{IR}} = \, a^4(y_{\rm IR}) \,\sqrt{ 1 - \frac{(\partial y_{\rm IR})^2}{a^2(y_{\rm IR})} }\,,
\ee 
where $(\partial y_{\rm IR})^2=\eta^{\mu\nu}\partial_\mu y_{\rm IR} \, \partial_\nu y_{\rm IR}$.
 Second, the Gibbons-Hawking, proportional to the extrinsic curvature at the $y=y_{\rm IR}(x^\mu)$ surface, generates additional terms. The relevant ones (contributing to the quadratic kinetic part) are proportional to the derivative of the warp factor at the brane location.

At this point we must make a slight detour, to be more precise on how several quantities depend on $y_{\rm IR}$, that is, on the dilaton. The key point is that the bulk scalar $\phi$  is coupled to the IR-brane (because the IR potential $V_b(\phi)$  acts effectively like a scalar charge). For this reason, the profile of $\phi$ (and therefore of the metric) in the bulk actually depends on the IR-brane location even when we allow the brane location to be off shell. To make this dependence manifest, we can write that the field profile is a function of both the bulk coordinate and the IR-brane location, $\phi=\phi(y,y_{\rm IR})$%
\footnote{We could also include the dependence on the UV brane location, $y_{\rm UV}$, however we will omit it here to avoid clutter. The symmetries of the background ensure that $\phi=\phi(y-y_{\rm UV},y_{\rm IR}-y_{\rm UV})$. This makes manifest that the field evaluated at the IR brane, $\phi(y_{\rm IR}-y_{\rm UV},y_{\rm IR}-y_{\rm UV})$,  is sensitive to $y_{\rm UV}$.}
. 
This is indeed implied by \eq{eqtachyon} and \eq{phit} in the main text.
In this notation, the field evaluated on the IR-brane is $\phi(y_{\rm IR},y_{\rm IR})$, and the derivative with respect to $y_{\rm IR}$ originates from the two arguments. The boundary condition specifies $\partial_y \phi(y,y_{\rm IR})|_{y=y_{\rm IR}}$, but the `full' derivative 
$\partial_{y_{\rm IR}} \phi(y_{\rm IR}, y_{\rm IR})$ is left unspecified and it is nontrivial in a nonlinear theory. As we will see shortly this full derivative is the one that controls the dilaton mass (it is the one that appears in \eq{md1}). 

The same qualifications apply also for the metric. After all, the `Friedman' \eq{A.B3} forces $\partial_y a$ to be an algebraic function of $\phi(y,y_{\rm IR})$ and $\partial_y \phi(y,y_{\rm IR})$, so the warp factor profile too depends parametrically on $y_{\rm IR}$, that is, we must write $a=a(y,y_{\rm IR})$. Now, the extrinsic curvature is related to $\partial_y a(y,y_{\rm IR}) |_{y=y_{\rm IR}}$. The dilaton variable $\pd$ defined in \eq{phiDy} stands for $a(y_{\rm IR},y_{\rm IR})$, and its derivative with respect to  $y_{\rm IR}$, $\partial_{y_{\rm IR}} a(y_{\rm IR})$  is not the same as $\partial_y a(y,y_{\rm IR}) |_{y=y_{\rm IR}}$. To make the distinction clear in the following we will keep this notation and show explicitly the difference.    
Form \eq{A.B3}  we have  
\be
-\ln a(y_{\rm IR},y_{\rm IR}) = A(y_{\rm IR},y_{\rm IR})=\int_{y_{\rm UV}}^{y_{\rm IR}} d\bar y \sqrt{\f{1}{L^2}+\frac{\hat\kappa^2L^2}{12} \left(\frac{\dot\phi^2(\bar y,y_{\rm IR})}{2}-V(\phi(\bar y,y_{\rm IR}))\right)}\,,
\ee
that differentiating with respect to $y_{\rm IR}$ leads to 
\be
\partial_{y_{\rm IR}} A(y_{\rm IR},y_{\rm IR}) = \dot A(y_{\rm IR},y_{\rm IR}) + \frac{\hat\kappa^2L^3}{24} \int_{y_{\rm UV}}^{y_{\rm IR}}d\bar y \frac{\dot\phi \,\partial_{y_{\rm IR}} \dot\phi - V'(\phi) \,\partial_{y_{\rm IR}} \phi  }{\sqrt{1+\frac{\hat\kappa^2L^4}{12} (\dot\phi^2/2-V(\phi))}}~,
\ee
where $ \dot A (y_{\rm IR},y_{\rm IR}) \equiv ( \partial_{y} \ln{ a(y,y_{\rm IR})} ) \big|_{y=y_{\rm IR}} $ and the rest of the notation should be clear. This shows that both at small and large $\hat\kappa^2$, the difference between $\dot A$ and $\partial_{y_{\rm IR}} A$ is small.  Numerically, in our solutions we finds it to be less than 5\%. (The same cannot be said about the two types of derivatives acting on $\phi$.) 
%

Returning to the kinetic term: after collecting all terms and using the EOM of the background, one arrives at \cite{Garriga:2001ar}
\be\label{kin}
3 \frac{M_5}{\kappa^2}\,  \dot A(y_{\rm IR}) \, a^2(y_{\rm IR})  \, (\partial_\mu y_{\rm IR})^2\,.
\ee 
With the definition \eq{phiDy} that  implies $\partial_\mu \pd = - \partial_{y_{\rm IR}}\ln\,a(y_{\rm IR})\; \pd\;\partial_\mu y_{\rm IR}$, this gives
\be\label{kin2}
\frac{3  M_5 \,L^2 }{\kappa^2} \, \frac{\dot A(y_{\rm IR}) }{[ \partial_{y_{\rm IR}}A(y_{\rm IR}) ]^2 } \, 
(\partial_\mu\pd)^2\,.
\ee
As discussed above, one can set $\dot A(y_{\rm IR})  \simeq  \partial_{y_{\rm IR}}A(y_{\rm IR})$ to a good approximation, therefore  the kinetic term is to a good accuracy
\be\label{kin3}
\frac{3  M_5 \,L^2 }{\kappa^2} \, \frac{(\partial_\mu\pd)^2}{\partial_{y_{\rm IR}}A(y_{\rm IR})} \, .
\ee

On the other hand, differentiating \eq{minV2}  we get
\be
\frac{d^2V_{\rm eff}(\pd)}{d\pd^2}=-
N_F \,  L^4 \, M_5 \, \pd^{\,2} \left[
\frac{1}{\dot A} \left(\frac{V_b'\,V_b''}{M_5^2}-V'\right)+\frac{4}{M_5}\,V_b'
 \right]
 \; \frac{\partial_{y_{\rm IR}}\phi(y_{\rm IR})}{\partial_{y_{\rm IR}}A(y_{\rm IR})}\,,
\ee
by using the chain rule and Eq.~\eqref{phiDy}.
Taking everything together, and the approximate expression \eq{kin3} we find that  the physical dilaton mass is  given by
\be\label{mdil}
m^2_{\phi_d}\simeq -
\frac{\hat\kappa^2 L^4}{ 6 }
\, \pd^{\, 2} 
\left[ \frac{1}{\dot A}  \left(\frac{V_b'\,V_b''}{M_5^2}-V'\right)+\frac{4}{M_5}\,V_b'
 \right]
 \;  \partial_{y_{\rm IR}}\phi(y_{\rm IR})\,, 
\ee
where everything is evaluated at the minimum, 
which coincides with \eq{md1} after the change of coordinates $dy= a(z) dz$.
As a further cross-check, let us note that this expression agrees with Eq.~(2.25) of \cite{Megias:2015ory}. This was obtained starting directly from \eq{A.B6} and obtaining an expression for the lowest KK mass under the assumption that it is light. 
The formula of \cite{Megias:2015ory} has  two distinct limits corresponding to whether or not the light dilaton is incarnated by the IR-brane position.
It is easy to check that in the limit where  the dilaton is the displacement of the IR-brane,  \eq{mdil} agrees with Eq.~(2.25) of  \cite{Megias:2015ory}.

Finally, we remark that using the EOM of the background we can rewrite \eq{kin} as
\be\label{kin4}
- \frac12\, V_b\big(\phi(y_{\rm IR})\big) \; a^2(y_{\rm IR}) \; (\partial_\mu y_{\rm IR})^2 ~.
\ee
One immediately realizes an important implication: the positivity of the kinetic energy restricts $V_b(\phi(y_{\rm IR}))<0$.
This is equivalent to restricting the effective tension on the IR-brane to be negative -- as it should be in order that it gives an end to the geometry at $y_{\rm IR}$. 
If we demand that this constraint is satisfied by all the solutions, including the one with  $\phi(z)=0$ (for $\epsilon<0$), then this translates into a constraint on 
the IR-brane tension `detuning 
parameter',
\be\label{upperbound}
\delta\hat\Lambda < 6~.
\ee
This reproduces the upper bound in \eqref{bounds}, {\em i.e.}, that the brane on the IR (on the `interior') side of the geometry has negative tension.

%
%
%

\section{A tale of two scalars: the 4D effective potential of a tachyon and a dilaton}
\label{appB}

When  both 4D tachyon  and  dilaton masses are smaller than $1/\zir$, we can 
 easily understand the physics of the system by just looking at the 4D effective theory for these two modes.
This is possible  when working close to the critical point, $\zir\approx\zirc$ (limiting case I of Section~\ref{tasol}), 
where we can obtain the masses and quartic couplings of the model as a perturbation of the model around $\zir=\zirc$. 
We find 
\be
\frac{1}{M_5LN_F}V_{\rm eff}(\phi_t,\phi_d)= \frac{1}{2}\hat m^2(\phi_d) \phi_t^2\phi_d^2+\frac{1}{4}\lambda_\ef \phi_t^4
+\frac{1}{4}\lambda_d \phi_d^4\,,
\label{pot}
\ee
where $\phi_d= 1/\zir$  is the dilaton and $\phi_t$ the 4D tachyon.
We are working here with non-canonically normalized fields:
\be
\frac{1}{M_5LN_F} \, {\cal L}_{\rm kin}=  \frac{1}{2}(\partial_\mu\phi_t)^2+\frac{3}{\hat\kappa^2}(\partial_\mu\phi_d)^2\,.
\ee  
\eq{pot} can only give a non-trivial minimum  
for $\phi_d\leq \mu_c\equiv1/\zirc$ such that  $\hat m^2(\phi_d)$  is negative.
In this regime, this is given by  
\be
\hat m^2(\phi_d)=\beta\ln({\phi_d}/{\mu_c})\,,
\label{m2val}
\ee
  where $\beta$ is given in \eq{tachy}.
Recall that  \eq{pot}  is only valid for $\hat m^2(\phi_d) \ll 1$ that requires either  $|\ln{\phi_d}/{\mu_c}|\ll 1$  or  $\beta\ll 1$ ($\hat m^2_b\to -2$).
The  tachyon quartic  $\lambda_\ef $ has a mild dependence on $\hat m_b^2$ and is derived below (Section~\ref{appQuartic}),
while the dilaton quartic is given by the detuning of the IR-brane tension: 
\be
\lambda_d = 4\frac{ \delta\hat\Lambda}{\hat \kappa^2}\,.
\label{lambd}
\ee
The presence of the tachyon leads to a non-trivial potential for the dilaton with a  minimum at
\be
\ln\frac{\langle\phi_d\rangle}{\mu_c}= -\frac{1}{4}\left[1+\sqrt{1+16\lambda_t\lambda_d/\beta^2}\,\right]\,.
\label{minpot}
\ee
We see that the largest  value of the dilaton is given by 
$\ln{\langle\phi_d}\rangle/{\mu_c}\leq-1/4$ that implies that the IR-brane can never be stabilized 
very close to the critical point $\zirc$ where the tachyon mass is small.
At the closest value $\ln{\langle\phi_d}\rangle/{\mu_c}=-1/4$, one finds that the minimum is  an inflection point
where the minimum coincides with a maximum (and one can check that the dilaton mass is zero at this point).
Nevertheless,  demanding   that the quartic couplings are positive, to guarantee that for $\epsilon<0$ the theory 
is conformal ($\phi_d\to 0$), one obtains $\ln{\langle\phi_d}\rangle/{\mu_c}\leq-1/2$.

\begin{figure}[t]
\centering
\hskip-.4cm
\includegraphics[width=.55\textwidth]{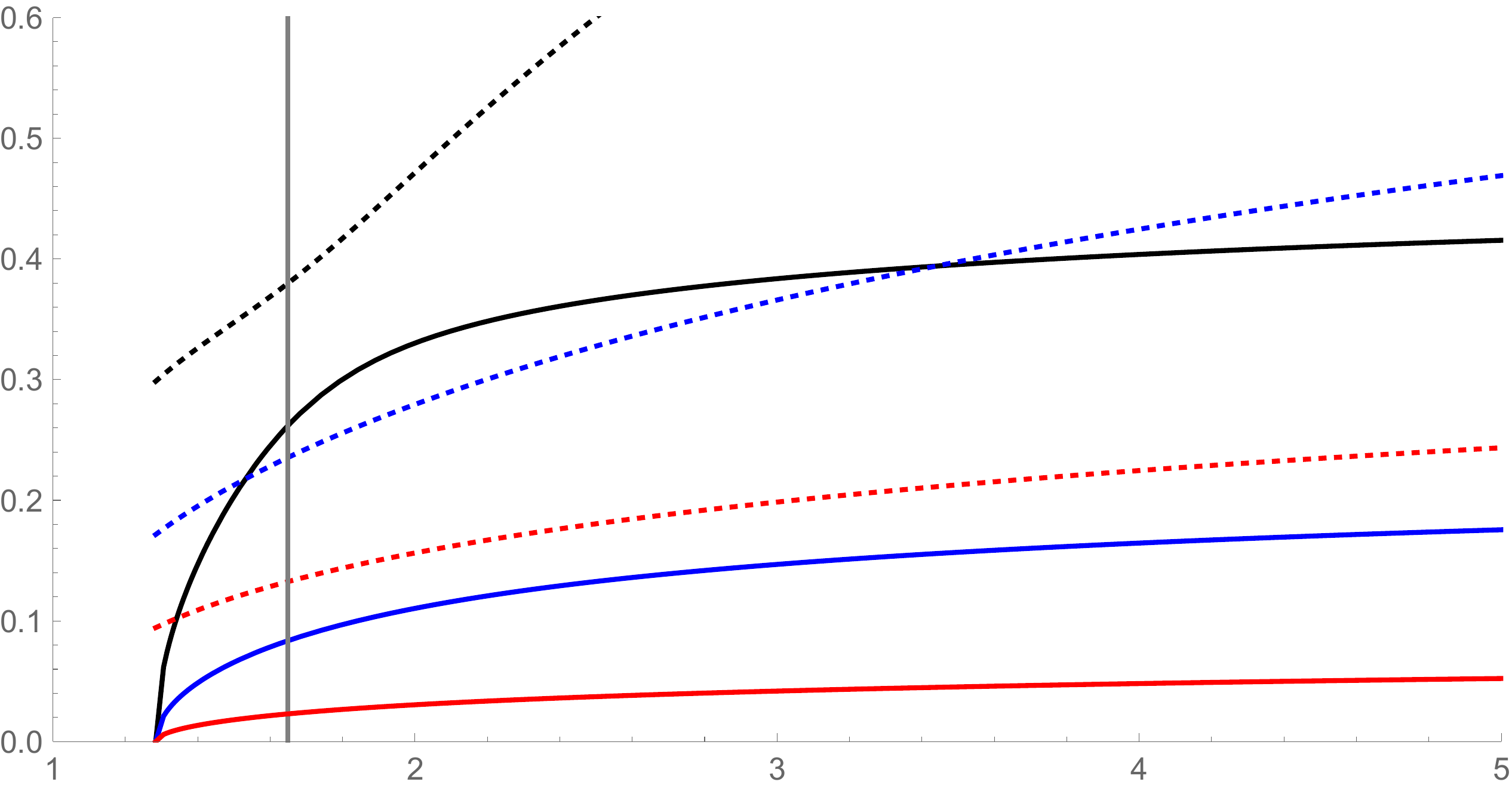}
\caption{\it 
Masses of the two scalars, that play the role of $S_1$ and $S_2$,  as a function of $\mu_c/\langle \phi_d \rangle$ and normalized to $(3\pi/4)\;\langle\phi_d\rangle$. We have taken $\lambda=1$, $\hat\kappa^2=6$  and  $\hat m_b^2=-0.5,\, -1.5,\,-1.75$ for the black, blue and red  lines respectively. The solid (dashed) line indicate  the lightest (heaviest) mode. 
The quartic coupling $\lambda_d$  varies along the horizontal axis according to \eq{minpot}.   
The vertical line marks where  $\lambda_d=0$, having $\lambda_d>0$ in the region to the right of it.}
\label{massdilaton2}
\end{figure}

We can also find the masses of the dilaton and tachyon by calculating the 
eigenvalues of the matrix of second derivatives at the  minimum \eq{minpot} after canonically 
normalizing the fields.
These give rather complicated functions of $\lambda_\ef$, $\lambda_d$ and $\beta$. 
For $\beta^2/(\lambda_d\lambda_\ef)\ll 1$ they reduce to
\be
m_{\phi_d}^2\simeq \frac{2\beta\,\hat\kappa^2}{\hat\kappa^2 +6\sqrt{\lambda_\ef/\lambda_d}}  \; \langle\phi_d\rangle^2\ ,\ \ \ \ 
m_{\phi_t}^2\simeq \lambda_d\left( {\frac{\hat\kappa^2}{6}+\sqrt{\lambda_\ef/\lambda_d} }\right) \; \langle\phi_d\rangle^2 \,,
\label{betasmall}
\ee
while for  $\beta^2/(\lambda_d\lambda_\ef)\gg 1$ these are
\be\label{mdil_lambda0}
m_{\phi_d}^2\simeq \frac{\hat\kappa^2}{6} \, \frac{ \beta^2 }{ 2\lambda_\ef} \; \langle\phi_d\rangle^2\ ,\ \ \ \ 
m_{\phi_t}^2\simeq \beta \; \langle\phi_d\rangle^2 \,.
\ee
With an abuse of notation, we have identified the lightest of the two modes as the dilaton, even though the eigenmodes corresponding to \eq{mdil_lambda0} can have a sizeable mixing in the $\phi_d-\phi_t$ basis.
We show the full  dependence on the parameters  in Fig.~\ref{massdilaton2}, where we keep $\lambda_\ef$ and $\beta$ fixed and vary $\lambda_d$ for various values of $\hat m_b^2$ (that is, of $\lambda_\ef$ and $\beta$). 
One clearly sees several features: 
\begin{enumerate}
\item[i)] the solutions  `start'  at $\ln(\mu_c/\langle\phi_d\rangle)=1/4$ where 
 the dilaton is  massless,  corresponding  to the inflection point.
 This requires however  $\lambda_d<0$  that we already said  is inapplicable.

\item[ii)]   the dilaton mass is suppressed by one power of $\beta$ for  $\lambda_t\lambda_d\sim O(1)$ (\eq{betasmall}). 
In our model $\beta$ can be small only near $\hat m_b^2=-2$. 
In that case the suppression reads $m^2_{\phi_d} \sim (\hat m_b^2+2)^2$. Keep in mind, however, that $\hat m_b^2=-2$ is not protected by any symmetry, so this is not representative of the full allowed parameter space. 

\item[iii)] the shapes of the lines resemble qualitatively those of Fig.~\ref{plotmassdilaton}.
Nevertheless, the agreement between this and the 5D model is only expected  for small $\hat m^2(\phi_d)$, since this measures how large is the tachyon  VEV.
A similar analysis can be done with more general choices of $\hat m^2(\phi_d)$ and the same qualitative behavior is observed quite generically as long as $\hat m^2(\phi_d)$ changes sign and has a moderate dependence on $\phi_d$. Interestingly enough, it suffices to take that $\hat  m^2(\phi_d)$ goes to a constant as $\phi_d\to0$, in order to obtain a dilaton mass with a rising-decreasing shape  as in Fig.~\ref{plotmassdilaton}
\end{enumerate}

When the dilaton is lighter than the tachyon, for example for $\beta\ll 1$,
we can alternatively  integrate out the tachyon from \eq{pot} and obtain \eq{potdil} with
\be
\lambda_{\rm eff}(\phi_d)=\frac{\lambda_d}{4}-\frac{\beta^2}{4\lambda_t}\ln^2({\phi_d}/{\mu_c})\,,
\label{lameff}
\ee
that tells us that the explicit breaking of scale invariance is logarithmic,
as expected from the dual theory  due to the presence of 
the  double-trace marginal operator ${\cal O}_g$. 
 \eq{potdil} with \eq{lameff}   leads to \eq{minpot} and to the dilaton mass of \eq{betasmall}.
As expected the dilaton mass is proportional to 
\be
\beta_{\lambda_{\rm eff}}({\langle\phi_d\rangle})=-\frac{\beta^2}{2\lambda_t}\ln({\langle\phi_d\rangle}/{\mu_c})\sim \beta\,,
\ee
where in the last equality we have used \eq{minpot} with $\beta\ll 1$.

\subsection{Effective quartic coupling for the tachyon}
\label{appQuartic}

The quartic self-coupling for the 4D tachyon can be obtained readily by plugging into the 5D potential quartic term the normalized profile of the 5D tachyon field near the condensation point and performing the integral over $z$. In the limit $\epsilon\to0$, $\zuv/\zirc\to0$ with $\sqrt\epsilon\,\ln\left(\zuv/\zirc\right)$ finite, one obtains
\be\label{quarticlambda}
\lambda_\ef = \left( \frac{3}{8}+\frac{9+2\hat m_b^2}{2(10+6 \hat m_b^2 + \hat m_b^4)^2} \right){\lambda}\,.
\ee
Even without any quartic self-coupling $\lambda$, the tachyon field experiences a stabilizing effect from its coupling to the metric. 
This is manifest in the background equation \eq{A.B2}, because the `friction' term which depends on $\phi$ itself, see \eq{A.B3}. More explicitly, the metric can be integrated out by using  \eq{A.B3} to obtain a closed equation for $\phi$
$$
\ddot{\phi}=4\,\sqrt{\f{1}{L^2}+\f{\hat\kappa^2L^2}{12}\Big(\f{\dot{\phi}^2}{2}-V(\phi)\Big)}\;\dot{\phi}+V'(\phi)\,.
$$
At leading order in $\hat\kappa^2$, one identifies a cubic term in the equation of motion
$$
\f{\hat\kappa^2L^3}{6}\Big(\f{\dot{\phi}^2}{2}-V(\phi)\Big)\;\dot{\phi}~.
$$
This suggests identifying the effective quartic coupling from the 5D integral of $\phi $ times the previous expression with the normalized tachyon profile. This gives
\be\label{quartick}
\Delta\lambda_\ef =\hat\kappa^2\, \frac{128+128 \hat m_b^2 +60 \hat m_b^4+12 \hat m_b^6+\hat m_b^8}{6(10+6 \hat m_b^2 + \hat m_b^4)^2}\,.
\ee
The expressions  in \eq{quarticlambda} and \eq{quartick} define  the functions $c_{\lambda,\kappa}$ introduced in \eq{effla}.

\end{document}